\documentclass[sigconf,10pt]{acmart}

\usepackage{silence}
\usepackage{siunitx}

\usepackage[english]{babel}
\usepackage{graphicx,booktabs,amsmath,times}
\usepackage{tabularx,multirow}
\usepackage{hyphenat,xspace,color,colortbl}
\usepackage{microtype}
\usepackage[labelformat=simple,skip=0pt]{subcaption}
\usepackage{mathtools}
\DeclarePairedDelimiter\ket{\lvert}{\rangle}
\usepackage{xcolor,enumitem}

\definecolor{Gray}{gray}{0.9}
\definecolor{LightGreen}{rgb}{0.88,1,0.88}
\definecolor{LightOrange}{rgb}{1,0.85,0.8}
\definecolor{LightRed}{rgb}{1,0.80,0.80}
\definecolor{LightBlue}{rgb}{0.8,0.8,1.0}

\renewcommand\thefigure{\arabic{figure}}
\renewcommand\thetable{\arabic{table}}

\renewcommand\footnotetextcopyrightpermission[1]{} 
\setcopyright{none}
\acmDOI{}
\acmISBN{}
\acmPrice{}
\newsavebox{\measurebox}

\newcommand{\parahead}[1]{\vspace{2pt plus 0pt minus 2pt}\noindent{\itshape #1}}
\newcommand{\parabreak}{\vspace*{1.00ex minus 0.25ex}\noindent}

\renewcommand{\paragraph}[1]{\vspace{2pt plus 0pt minus 2pt}\noindent{\bfseries #1}}

\begin{document}
\title{A Cost and Power Feasibility Analysis of Quantum\\
Annealing for NextG Cellular Wireless Networks}

\author{Srikar Kasi$^{\star,\dagger}$, P.A.Warburton$^\ddagger$, John Kaewell$^\dagger$, Kyle Jamieson$^\star$}
\affiliation{\vspace{-9pt} \small $^\star$Princeton University, $^\dagger$InterDigital, Inc., $^\ddagger$University College London}

\begin{abstract}
In order to meet mobile cellular users' ever-increasing data demands, today's 4G and 5G networks are designed mainly with the goal of maximizing spectral efficiency. While they have made progress in this regard, controlling the carbon footprint and operational costs of such networks remains a long-standing problem among network designers. This paper takes a long view on this problem, envisioning a NextG scenario where the network leverages quantum annealing for cellular baseband processing. We gather and synthesize insights on power consumption, computational throughput and latency, spectral efficiency, operational cost, and feasibility timelines surrounding quantum technology. Armed with these data, we analyze and project the quantitative performance targets future quantum annealing hardware must meet in order to provide a computational and power advantage over CMOS hardware, while matching its whole-network spectral efficiency. Our quantitative analysis predicts that with quantum annealing hardware operating at a 102~$\mu$s problem latency and 3.1M qubits, quantum annealing will achieve a spectral efficiency equal to CMOS computation while reducing power consumption by 41 kW (45\% lower) in a representative 5G base station scenario with 400~MHz bandwidth and 64 antennas, and an 8~kW power reduction (16\% lower) using 1.5M qubits in a 200~MHz-bandwidth 5G scenario.
\end{abstract}

\maketitle

\section{Introduction}
\label{s:intro}

Today's 4G and 5G Cellular Radio Access Networks (RANs) are experiencing unprecedented
growth in traffic at base stations (BSs) due to increased subscriber numbers 
and their higher quality of service requirements \cite{cisco, nokia}. To meet the resulting
demand, techniques such as Massive Multiple-Input Multiple-Output (MIMO) communication, 
cell densification, and millimeter\hyp{}wave communication are expected to 
be deployed in fifth\hyp{}generation (5G) cellular standards \cite{3gppintro}. But this
in turn significantly increases the power and cost required to operate RAN sites
backed by complementary metal oxide semiconductor (CMOS)-based computation. While research and industry efforts have 
provided general solutions (\textit{e.g.,} sleep mode \cite{lahdekorpi2017energy} and 
network planning \cite{wu2015energy}) to increase energy efficiency and decrease power 
consumption of RANs, the fundamental challenge of power requirements scaling 
with the exponentially increasing computational requirements of the RAN persists. 
Previously (\emph{ca.}~2010), this problem had not limited  
innovation in the design of wireless networks,
due to a rapid pace of improvement in CMOS's computational efficiency.
Unfortunately however, today, such developments are not maintaining the pace
they had in past years, due to transistors approaching atomic limits 
\cite{courtland2016transistors} and the end of Moore's Law 
(expected \emph{ca.}~2025--2030 \cite{khan2018science, shalf2020future, itrs}). 
This therefore calls into question the prospects of CMOS to
achieve NextG cellular targets in terms of both energy and spectral efficiency.
\begin{figure}
\centering
\includegraphics[width=0.75\linewidth]{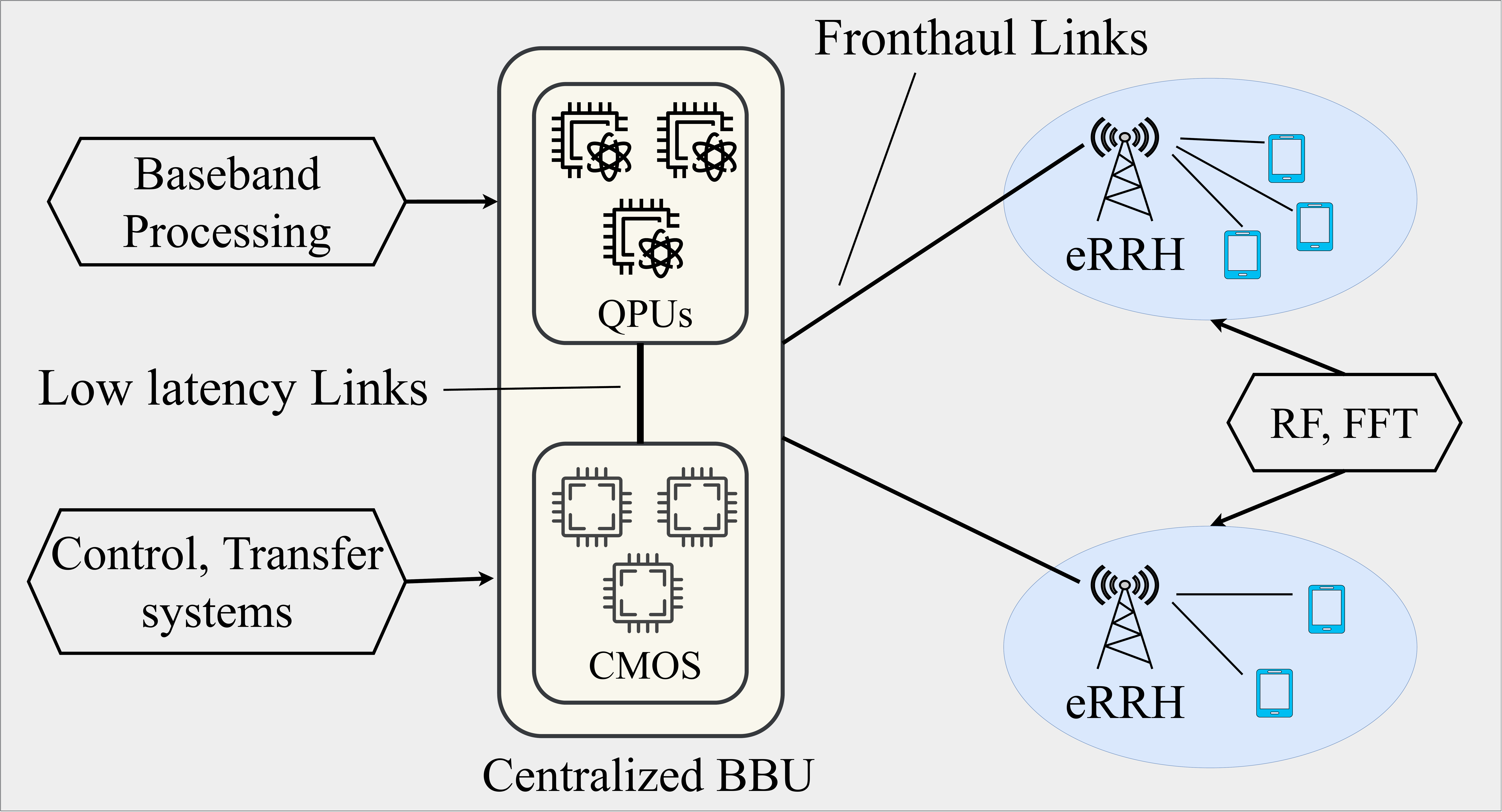}
\caption{Our envisioned
deployment scenario of Quantum Processing Units (QPUs) alongside CMOS units
in a C-RAN datacenter. QPUs undertake heavy baseband computation, while
CMOS processing manages the network's control plane.}
\label{fig:cran}
\end{figure} 
\begin{figure*}[ht]
\centering
\includegraphics[width=0.97\linewidth]{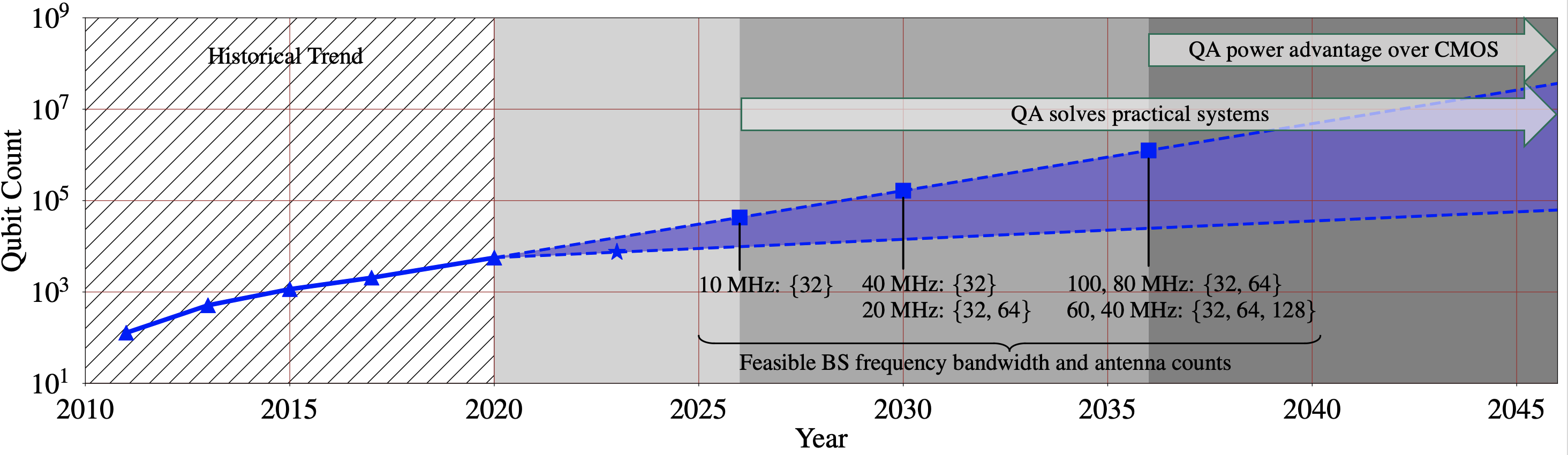}
\caption{Projected year-by-year timeline of QA-based RAN processing. Data points ($\begingroup\color{blue}\blacktriangle\endgroup$) in the hatched area (2011--2020) represent the historical QA qubit counts. The 2023 data point ($\begingroup\color{blue}\star\endgroup$) with 7,440 qubits corresponds to a next-generation QA processor roadmap \cite{clarity, zephr}. The blue filled (dark shade) area is the projected QA qubit count, whose upper/lower bounds are extrapolations of the best-case (2017--2020) and the worst-case (2020--2023) qubit growths respectively. Annotations corresponding to further data points ($\begingroup\color{blue}\blacksquare\endgroup$) show the base station (BS) scenarios their respective qubit counts will enable (see \S\ref{s:timelines}). The figure shows that if future QA qubit count scales along this best-case trend, starting from the year 2036, QA may be applicable to practical wireless systems with power/cost benefits over CMOS hardware (see \S\ref{s:timelines}).}
\label{fig:timeline}
\end{figure*}

\parabreak{}This work investigates a radically different baseband processing architecture 
for RANs, one based on quantum computation, to see whether this emerging technology 
can offer cost and power benefits over CMOS computation in wireless networks. 
We seek to quantitatively analyze whether in the coming years and decades, mobile operators 
might rationally invest in the RAN's capital (CapEx) by purchasing quantum 
hardware of high cost, in a bid to lower its operational expenditure (OpEx) and hence
the \emph{Total Cost of Ownership} (TCO~=~CapEx + OpEx). The OpEx cost reduction 
would result from the reduced power consumption of the RAN, due to higher
computational efficiency of quantum processing over CMOS processing for certain 
heavyweight baseband processing tasks.  Figure~\ref{fig:cran} depicts this 
envisioned scenario, where quantum processing units (QPUs) co\hyp{}exist 
with traditional CMOS processing at Centralized RAN (C-RAN) Baseband Units (BBUs)
\cite{checko2014cloud, 7414384}. QPUs will then be
used for the BBU's heavy baseband processing, 
whereas CMOS will handle the network's lightweight control plane processing
(\textit{e.g.,} resource allocation, 
communication control interface), transfer systems (\textit{e.g.,} enhanced 
common public radio interface, mobility management entity), and further 
lightweight tasks such as pre\hyp{} and post\hyp{}processing
QPU\hyp{}specific computation.

This paper presents the first extensive analysis on power consumption and
quantum annealing (QA) architecture to make the case for the future
feasibility of quantum processing based RANs. While recent successful point\hyp{}solutions 
that apply QA to a variety of wireless network applications 
\cite{10.1145/3372224.3419207, 9500557, kim2019leveraging, 
wang2016quantum, cao2019node, ishizaki2019computational, wang2019simulated, 
chancellor2016direct, lackey2018belief, bian2014discrete, cui2021quantum} serve as our motivation,
previous work stops short of a holistic power and cost comparison between
QA and CMOS. Despite QA's benefits demonstrated by these prior works in their 
respective point settings, a reasoning of how these results will factor 
into the overall computational performance and power requirements of the
base station and C-RAN remains lacking. Therefore, here we investigate these
issues head\hyp{}on, to make an end\hyp{}to\hyp{}end case that QA will likely
offer benefits over CMOS for handling BBU processing, and to make
time predictions on when this benefit will be realized.  Specifically,
we present informed answers to the following questions:
\begin{enumerate}[label={\bf Question \arabic*:},labelwidth=5.5em,itemindent=0em,leftmargin=5.5em,itemsep=3pt] 
    \item How many qubits (quantum bits) are required to realize BS or C-RAN BBU processing requirements? (\textbf{Answer:}~\emph{cf.}~\S\ref{s:qubitcount})
    \item Relative to purely CMOS BBU processing, how much power and cost does one save with such amount of qubits, viewed over the entire RAN? (\textbf{Answer:}~\emph{cf.}~\S\ref{s:results})
    \item At what year will these qubit numbers become feasible, based on recent prior trends in the industry? (\textbf{Answer:}~\emph{cf.}~\S\ref{s:intro}, \S\ref{s:timelines})
    \item To what amount will QA processing latency (a performance bottleneck in current QA devices) be reduced in the future? (\textbf{Answer:}~\emph{cf.}~\S\ref{s:latency})
\end{enumerate}

In order to realize the architecture of Figure~\ref{fig:cran},
several key system performance metrics need to be analyzed, quantified, and 
evaluated, most notably the computational throughput and latency (\S\ref{s:latency}), the power consumption of the entire system and
resulting
spectral efficiency (bits per second per Hertz of frequency spectrum)
and operational cost (\S\ref{s:results}).
Our approach is to first describe the factors that influence processing latency and
throughput on current QA devices and then, by assessing recent developments 
in the area, project what computational throughput and latency 
future QA devices will achieve
(\S\ref{s:latency}). We analyze cost by evaluating the power consumption of 
QA and CMOS\hyp{}based processing at equal spectral efficiency targets (\S\ref{s:results}). Our analysis reveals that a three\hyp{}way 
interplay between latency, power consumption, and the number of qubits available in the QA hardware determines whether QA can 
benefit over CMOS. In particular, latency influences spectral 
efficiency, power consumption influences energy efficiency, and the number
of qubits influences both. Based on these insights, we determine properties 
(\textit{i.e.,} latency, power consumption, and qubit count) 
that QA hardware must meet in order to provide an advantage over CMOS 
in terms of energy, cost, and spectral efficiency in wireless networks.

\begin{table}
\caption{Summary of qubit requirements of QA hardware to achieve equal 
spectral efficiency to CMOS, and power consumption of CMOS and
QA, at various bandwidths (B/W).\protect\footnotemark{} The shaded/colored cells indicate the lesser of the two power requirements of CMOS and QA.}
\label{t:powerbenefit}
\begin{small}
\begin{tabularx}{\linewidth}{*{7}{X}}\toprule
\multirow{3}{*}{\textbf{B/W}}& \multicolumn{2}{c}{\textbf{Qubits}}& 
    \multicolumn{4}{c}{\textbf{Power Consumption}}\\ \cline{2-3}\cline{4-7} \noalign{\smallskip}
& BS & CRAN &\multicolumn{2}{c}{BS (KW)}& \multicolumn{2}{c}{CRAN (MW)}\\ \cline{4-7}
\centering & & & CMOS& QA& CMOS& QA\\ 
\bottomrule \noalign{\smallskip} 
50~MHz  & 386K & 1.16M & \cellcolor{LightGreen}19.3 & 36 & \cellcolor{LightGreen}0.079 & 0.081\\
100 & 772K & 2.32M& \cellcolor{LightGreen}29.4 & 37.9 & 0.11 & \cellcolor{LightGreen}0.09\\
200 & 1.54M & 4.62M& 49.5 & \cellcolor{LightGreen}41.6 & 0.17 & \cellcolor{LightGreen}0.10\\
400 & 3.08M & 9.24M& 89.9 & \cellcolor{LightGreen}49 & 0.29 & \cellcolor{LightGreen}0.13\\
\bottomrule
\end{tabularx}
\end{small}
\end{table}

Table~\ref{t:powerbenefit} summarizes our results, 
showing that for 200 and 400~MHz bandwidths, respectively, 
with 1.54 and 3.08M qubits, we predict that 
QA processing will achieve spectral efficiency equal to today's 
14~nm CMOS processing, while reducing power consumption by 
8~kW (16\% lower) and 41~kW (45\% lower) in representative 5G base 
station scenarios. 
In a C-RAN setting with three base stations of 200 and 
400~MHz bandwidths, QA processing with 4.62M and 9.24M qubits, 
respectively, reduces power consumption by 70~kW (41\% lower) and
160~kW (55\% lower), while achieving equal spectral 
efficiency to CMOS.

Our further evaluations compare QA against future 
1.5~nm CMOS process, which is expected to be the silicon technology at 
the end of Moore's law scaling (\emph{ca.}~2030 \cite{itrs}). In a 
base station scenario with 400~MHz bandwidth and 128 antennas, QA with 6.2M\hyp{}qubits will reduce 
power consumption by 30.4~kW (37\% lower), in comparison to 1.5~nm CMOS,
while achieving equal spectral efficiency to CMOS.

Figure~\ref{fig:timeline} reports our
projected QA feasibility timeline, describing year-by-year milestones 
on the application of QA to wireless networks. Our analysis 
shows that with custom QA hardware (\textit{cf.}~\S\ref{s:assumptions}) and qubits growing 
2.65$\times$ every three years (the 2017--2020 trendline), 
QA application in practical RAN settings with potential power/cost benefit 
is a predicted 15 years (\textit{ca.} 2036) away, whereas the 
feasibility in processing for a base station (BS) 
with 10~MHz bandwidth and 32 antennas is a predicted 
five years away (\textit{ca.} 2026)
(\emph{cf.}~\S\ref{s:timelines}).

Overall, our quantitative results predict that QA hardware will 
offer power benefits over CMOS hardware in certain wireless network scenarios,
once QA 
hardware scales to at least a million qubits (\textit{cf.}~\S\ref{s:timelines}) and 
reduces its problem processing time to hundreds of 
microseconds, which we argue is feasible within our projected timelines. 
Scaling QA processors to millions of qubits will pose challenges related to
engineering, 
control, and operation of hardware resources, which designers
continue to investigate \cite{boothby2021architectural, bunyk-architectural}. Recent further
work
demonstrates large-scale qubit control 
techniques, showing that control of million qubit\hyp{}scale quantum hardware 
is already at this point in time a realistic prospect~\cite{vahapoglu2021single}.

\paragraph{Roadmap.} In the remainder of this paper,  Section~\ref{s:assumptions} describes background and assumptions,
Section~\ref{s:latency} analyzes QA hardware architecture and its end-to-end processing latency, and Section~\ref{s:model} describes power modeling in RANs and cellular computational targets. We will then be in a position to present our CMOS versus QA power comparison methodology and results in Section~\ref{s:results}. We conclude by discussing a projected feasibility timeline of QA-based RANs in Section \ref{s:timelines}.

\footnotetext{Silicon results reflect 14~nm CMOS process; QA results
reflect 102~$\mu$s problem processing latency (\emph{cf.}~\S\ref{s:latency}).
System parameters correspond to a base station with 64-antennas, 
64-QAM modulation, 0.5 coding rate, and 100\% time and frequency domain duty cycles. C-RAN handles three base stations.}

\section{Background and Assumptions}
\label{s:assumptions}

While classical computation uses bits  to process information,
quantum computation uses \textit{qubits}, physical devices that allow \textit{superposition} of bits simultaneously
\cite{tech}. The current technology landscape 
consists broadly of fault-tolerant approaches to quantum computing versus 
\textit{noisy intermediate scale quantum (NISQ)} implementations. 
Fault-tolerant quantum computing \cite{shor1996fault, preskill1998fault} is an ideal 
scenario that is still far off in the future, whereas 
NISQ computing \cite{preskill2018quantum}, which is 
available today, suffers high machine noise levels, but gives
us an insight into what future fault\hyp{}tolerant methods will be 
capable of in terms of key quantum effects such as
qubit \textit{entanglement} and \textit{tunneling} \cite{preskill2018quantum}. 
NISQ processors can be classified into digital 
gate model or analog annealing (QA) architectures.

Gate\hyp{}model devices \cite{knill2005quantum} are fully general purpose
computers, using programmable logic gates acting on qubits \cite{wichert2020principles}, whereas 
annealing\hyp{}model devices \cite{tech}, inspired by 
the Adiabatic Theorem of quantum mechanics, 
offer a means to search an optimization problem for
its lowest \emph{ground state} energy configurations 
in a high\hyp{}dimensional energy landscape \cite{boixo2013quantum}. While gate\hyp{}model quantum devices of size relevant to practical applications are not yet generally available \cite{ibm}, today's QA devices with about 5,000 qubits enable us to commence empirical studies at realistic scales \cite{tech}. Therefore we conduct this study from the perspective of annealing-model devices.

\subsection{Quantum Annealer Hardware} 
\begin{figure}
\centering
    
    \begin{subfigure}[b]{0.5\linewidth}
        \centering
        \includegraphics[viewport= 310 300 470 460,clip,width=0.73\linewidth]{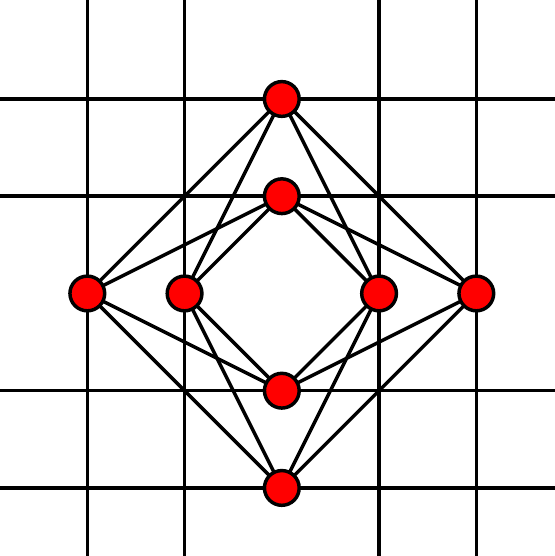}
        
        \caption{Chimera Unit Cell}
        \label{f:chimera}
    \end{subfigure}\hfill
    \begin{subfigure}[b]{0.5\linewidth}
        \centering
        \includegraphics[viewport= 810 345 910 445,clip,width=0.73\linewidth]{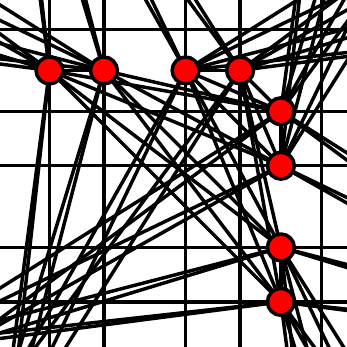}

        \caption{Pegasus Unit Cell}
    \end{subfigure}
    
\caption{The figure shows \textit{unit cell} structures of \textbf{(a)} Chimera and \textbf{(b)} Pegasus QA hardware topologies. Nodes in the figure are physical qubits, and edges are physical couplers.}
\label{f:unitcells}
\end{figure}

Quantum Annealing (QA) is an optimization\hyp{}based approach that aims to find the lowest energy \emph{spin configuration} (\emph{i.e.}, solution)
of an \emph{Ising model} (defined in \S\ref{s: algorithm})
described by the time-dependent energy functional
(Hamiltonian):
\begin{equation}
    H(s) = - \Gamma(s)H_I + L(s)H_P
    \label{eq:annealing}
\end{equation}
where $H_I$ is the \emph{initial Hamiltonian}, $H_P$ is the (input) 
\emph{problem Hamiltonian}, \textit{s} ($\in$ [0, 1]) is a non-decreasing function of time called an \textit{annealing schedule}, 
$\Gamma(s)$ and L(s) are energy scaling functions of the transverse and longitudinal fields in the annealer respectively. Essentially, $\Gamma(s)$ determines the probability of \textit{tunneling} during the annealing process, and $L(s)$ determines the probability of finding the ground state of the
input problem Hamiltonian $H_P$ \cite{tech}. The QA hardware is a network of locally interacting radio-frequency superconducting qubits, organized in groups of \textit{unit cells}. Fig.~\ref{f:unitcells} shows the unit cell structures of recent (Chimera) and state-of-the-art (Pegasus) QA devices. The nodes and edges in the figure are \textit{qubits} and \textit{couplers} respectively (detailed below) \cite{10.1145/3372224.3419207}.

\subsection{Quantum Annealing Algorithm}
\label{s: algorithm}

The process of optimizing a problem in the QA is called \textit{annealing}. Starting with a high transverse field (\textit{i.e.,} $\Gamma(0) >> L(0) \approx 0$), QA initializes the qubits in a pre-known ground state of the initial Hamiltonian $H_I$, then gradually interpolates this Hamiltonian over time---decreasing $\Gamma(s)$ and increasing L(s)---by adiabatically introducing quantum fluctuations in a low-temperature environment, until the transverse field diminishes (\textit{i.e.,} $L(1) \gg \Gamma(1) \approx 0$). This time-dependent interpolation of the Hamiltonian is essentially the annealing algorithm. The Adiabatic Theorem then ensures that by interpolating the Hamiltonian slowly\footnote{If the adiabatic evolution is infinitely slow, then the annealing algorithm is guaranteed to find the global minima of $H_P$ \cite{smolin2014classical}.} enough, the system remains in the ground state of the interpolating Hamiltonian \cite{baldassi2018efficiency}. Thus during the annealing process, the system ideally stays in the local minima and probabilistically reaches the global minima of the problem Hamiltonian $H_P$ at its conclusion \cite{tech}.

The initial Hamiltonian takes the form $H_I$ = $\sum_i \sigma_i^x$, where $\sigma_i^x$ is the result of the \emph{Pauli-X} matrix $\big(\begin{smallmatrix}
  0 & 1\\
  1 & 0
\end{smallmatrix}\big)$ acting on the $i^{th}$ qubit. Thus, the initial state of the system is the ground state of this $H_I$, where each qubit is in an equal \textit{superposition} state $\frac{1}{\sqrt{2}} \left(\ket{-1} + \ket{+1}\right)$. The problem Hamiltonian is described by $H_P = \sum_i h_{i}\sigma_i^z + \sum_{i<j}J_{ij}\sigma_i^z\sigma_j^z$, where $\sigma_i^z$ is the result of the \textit{Pauli-Z} matrix $\big(\begin{smallmatrix}
  1 & 0\\
  0 & -1
\end{smallmatrix}\big)$ acting on the $i^{th}$ qubit, $h_i$ and $J_{ij}$ are the optimization problem inputs that the user supplies \cite{tech, amin2015searching}.

\parahead{Input Problem Forms.} QAs optimize Ising model problems, whose problem format matches the above problem Hamiltonian: $E = \sum_i h_{i}q_i + \sum_{i<j}J_{ij}q_iq_j$, where \textit{E} is the energy of the candidate solution, $q_i$ is the $i^{th}$ solution variable which can take on values in \{$-1, +1$\}, $h_i$ and $J_{ij}$ are called the \textit{bias} of $q_i$ and the \textit{coupling strength} between $q_i$ and $q_j$, respectively. Biases represent individual preferences of qubits to take on a particular classical value ($-1$ or $+1$), whereas coupling strengths represent pairwise preferences (\textit{i.e.,} two particular qubits should take on same/opposite values), in the solution the machine outputs. Biases and coupling strengths are specified to qubits and couplers, respectively, using a programmable on-chip control circuitry \cite{johnson2011quantum, king2018observation}. The QA returns the solution variable configuration with the minimum energy \textit{E} at its output \cite{10.1145/3372224.3419207}.

\paragraph{Assumption 1--- Ising Model formulation.}
To enable QA computation, cellular baseband's 
heavy processing tasks must be formulated as Ising model problems.
Recent prior work in this area has formulated 
the most heavyweight tasks in the baseband, such as 
frequency domain detection, forward error correction, and 
precoding problems, into Ising models \cite{kim2019leveraging, 10.1145/3372224.3419207, 9500557, bian2014discrete, cui2021quantum}. Further baseband tasks (\emph{e.g.} filtering) will either admit
Ising model formulations via binary representation
of continuous values \cite{aschbacher2005digital, mattingley2010real} (we leave for future work), or 
are so lightweight they require negligible power.

\begin{figure}
\centering
\includegraphics[width=0.85\linewidth]{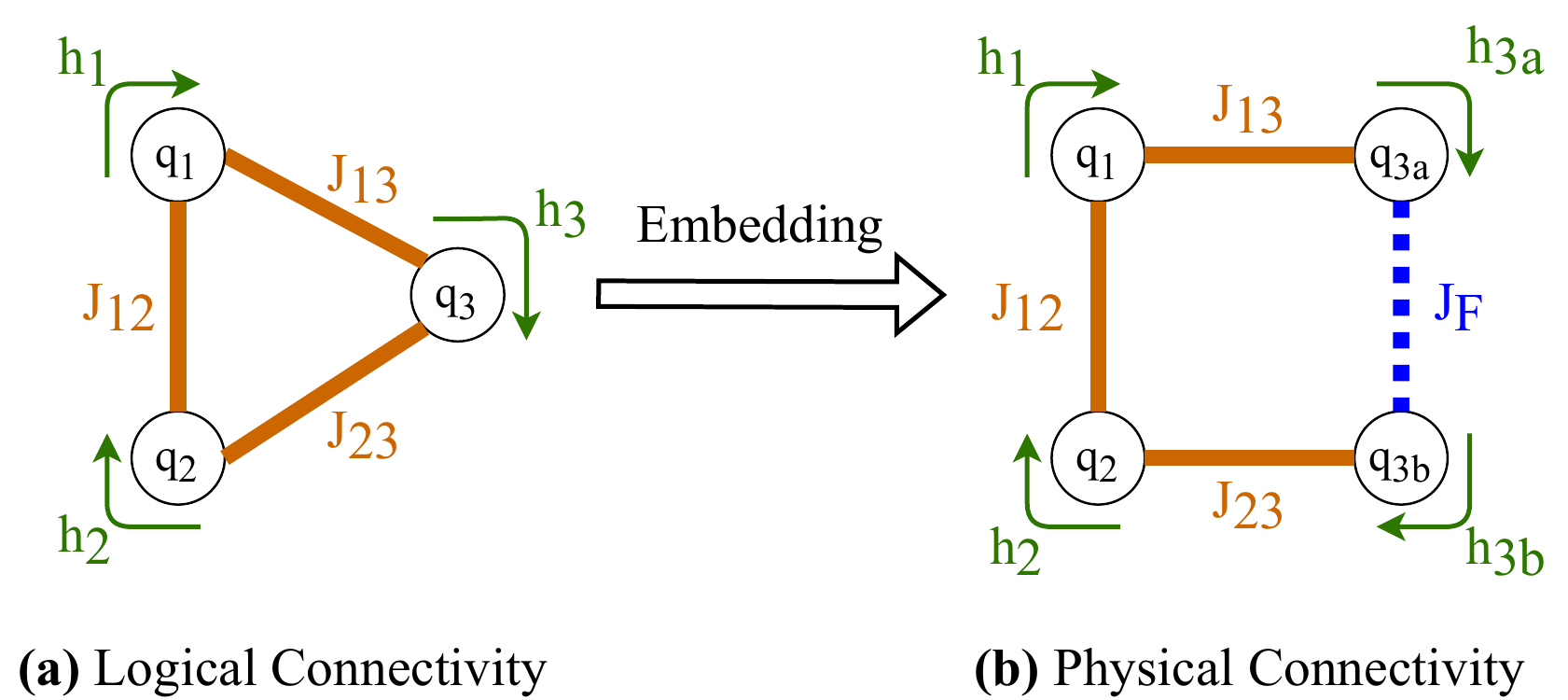}
\caption{The figure shows embedding process of Eq.~\ref{eq:embedding}, where the logical variable $q_3$ in (a) is mapped onto two physical qubits $q_{3a}$ and $q_{3b}$ as in (b) with a JFerro of $J_F$ (dotted).}
\label{fig:example}
\end{figure}
\begin{figure*}
\centering
\includegraphics[width=0.8\linewidth]{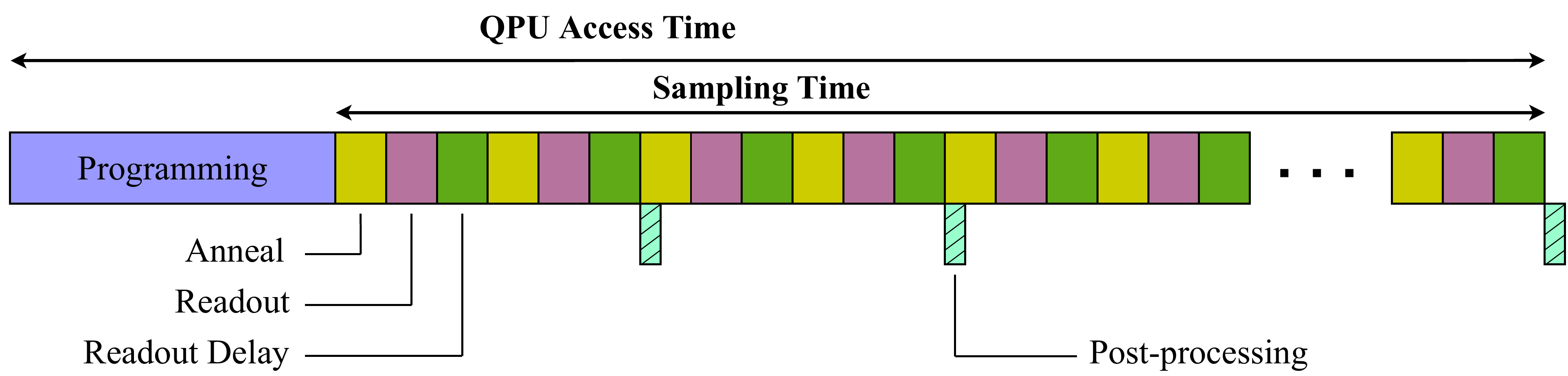}
\caption[Timing diagram]{Timing diagram of a quantum annealer device.  Machine access overheads 
not relevant to our proposed use case are omitted. Post-processing runs on integrated silicon, in parallel with the annealer computation \cite{tech}.}
\label{fig:timing}
\end{figure*} 
\subsection{Input Problem Embedding} The process of mapping a given input problem onto the physical QA hardware is called \textit{embedding}. To understand embedding, let us consider an example Ising problem:
\begin{equation}
    E = h_1q_1 + h_2q_2 + h_3q_3 + J_{12}q_1q_2 + J_{23}q_2q_3 + J_{13}q_1q_3
    \label{eq:embedding}
\end{equation}
The direct/logical representation of Eq.~\ref{eq:embedding} is depicted in Fig.~\ref{fig:example}(a), where nodes and edges in the figure are qubits and couplers respectively. The curved arrows in the figure are used to visualize the linear coefficients. However, observe that a complete three-node qubit connectivity does not exist in the Chimera graph (\textit{cf.} Fig.~\ref{f:unitcells}(a)). Hence the standard approach is to map one of the logical problem variables (\textit{e.g.,} $q_3$) onto two physical qubits (\textit{e.g.,} $q_{3a}$ and $q_{3b}$) as Fig.~\ref{fig:example}(b) shows, such that the resulting connectivity can be realized on the QA hardware.

To ensure proper embedding: $q_{3a}$ and $q_{3b}$ must agree with each other. This is achieved by enforcing the condition $h_3 = h_{3a} + h_{3b}$, and chaining these physical qubits with a strong ferromagnetic coupling called \textit{JFerro} ($J_F$)---see dotted line in Fig.~\ref{fig:example}(b). The physical Ising problem the QA optimizes for the example in Eq.~\ref{eq:embedding} is then:
\begin{align}
\begin{split}
E = h_1q_1 + h_2q_2 + h_{3a}q_{3a} + h_{3b}q_{3b} + J_{12}q_1q_2 + \\
J_{13}q_1q_{3a} + J_{23}q_2q_{3b} + J_Fq_{3a}q_{3b}
\end{split}
\label{eqn: example_phy}
\end{align}

\paragraph{Assumption 2--- Bespoke QA hardware}. Qubit connectivity
significantly impacts performance, with
sparse connectivity negatively affecting dense problem graphs
due to problem mapping difficulties \cite{10.1145/3372224.3419207}. Recent
advances in QA have bolstered qubit connectivity---6 to 15 to 20 couplers per qubit in the Chimera (2017), Pegasus (2020), and Zephyr (\textit{ca.} 2023-24) topologies respectively \cite{topologies, zephr}---while further improvement efforts 
continue \cite{katzgraber2018small, lechner2015quantum}, 
which will allow QA hardware tailored to baseband processing problems within
the timescales of our predictions, resulting in a highly efficient minor embedding process.

\section{Quantum Processing Performance}
\label{s:latency}

To characterize current and future QA performance, this section analyzes processing time on QA devices, the client of which
sends \textit{quantum machine instructions} (QMI) that characterize an 
input problem computation to a QA QPU. The QPU then responds with solution data. Fig.~\ref{fig:timing} depicts the the entire latency 
a QMI experiences from entering the QPU to the readout of the solution, 
which consists of \textit{programming} (\S\ref{s:prog}), \textit{sampling} (\S\ref{s:sampling}), 
and \textit{post\hyp{}processing} (\S\ref{s:post}) times.

\subsection{Programming} \label{s:prog} As the QMI reaches the QPU, the QPU programs the QMI's 
input problem coefficients---biases and coupling strengths (\S\ref{s:assumptions}):
room temperature electronics send raw signals into the QA 
refrigeration unit to program the on\hyp{}chip flux 
digital\hyp{}to\hyp{}analog converters (\emph{$\Phi$-DACs}). 
The $\Phi$-DACs then apply external magnetic fields
and magnetic couplings locally to the qubits and 
couplers respectively. This process is called a 
\textit{programming cycle}, and in current technology it 
typically takes 4--40~$\mu$s~\cite{timing},
dictated by the bandwidth of control lines and 
the $\Phi$-DAC addressing scheme \cite{boothby2021architectural}.
During the programming cycle, the QPU dissipates an amount of heat 
that increases the effective temperature of the qubits. This is due to 
the movement of flux quanta\footnote{QA devices store coefficient 
information in the form of magnetic flux quanta and it is transferred 
via single flux quantum (SFQ) voltage pulses \cite{bunyk-architectural}.} 
in the inductive storage loops of $\Phi$-DACs. Thus, a 
post\hyp{}programming \textit{thermalization} time is required to
cool the QPU, ensure proper reset\fshyp{}initialization
of qubits, and allow the QPU to maintain a thermal equilibrium
with the refrigeration unit
($\approx$20~mK). QA clients can specify thermalization 
times in the range 0--10~ms with microsecond-level granularity. The default value
on D-Wave's machine is a conservative one millisecond~\cite{tech}.
\begin{table}
\caption{The QPU on-chip energy dissipation values for the worst-case programming (\textit{i.e.,} using all qubits and couplers) and their associated thermalization time required for various choices of QPU sizes and $\Phi$-DAC critical currents.}
\label{t:energy}
\begin{small}
\begin{tabularx}{\linewidth}{X*{5}{c}}\toprule
\centering \textbf{Qubits} & \textbf{Couplers} & \textbf{$\Phi$-DACs} & \multicolumn{2}{c}{\textbf{Energy, Thermalization time}}\\ \cline{4-5} \noalign{\smallskip}
\centering & & & $I_c$ = 55~$\mu$A \cite{bunyk-architectural} & $I_c$ = 1~$\mu$A \cite{mcdermott2018quantum} \\
\bottomrule \noalign{\smallskip} 
\centering 512& 1,472 & 4,544 & $\approx$ 66~fJ, 2.2~ns & $\approx$ 1~fJ, 33~ps \\
\centering 2,048& 6,016 & 18,304 & $\approx$ 266~fJ, 8.9~ns & $\approx$ 5~fJ, 167~ps\\
\centering 5,436& 37,440 & 70,056 & $\approx$ 1~pJ, 33~ns & $\approx$ 18~fJ, 600~ps\\
\centering 10~M& 75~M & 135~M & $\approx$ 2~nJ, 66~$\mu$s & $\approx$ 36~pJ, 1.2~$\mu$s\\
\bottomrule
\end{tabularx}
\end{small}
\end{table}
QMI coefficients are programmed by using six $\Phi$-DACs per qubit and 
one $\Phi$-DAC per coupler, and the supported bit\hyp{}precision is 
currently up to five bits (four for value, one for sign) 
\cite{bunyk-architectural}. Each $\Phi$-DAC consists two 
inductor storage loops with a pair of Josephson junctions each. 
The energy dissipated on chip is on the order of $I_c\times \Phi_0$ 
per \textit{single flux quantum (SFQ)} moved in an inductor storage loop, 
where $I_c$ is the $\Phi$-DAC's junction critical current and 
$\Phi_0$ is the magnetic flux quantum.\footnote{$\Phi_0 = h/2e$, where 
\textit{h} is Planck's constant and \textit{e} is the electron charge.} 
For the worst-case reprogramming scenario, this corresponds to 32 
SFQs ($-$16 to $+$16) moving into (or out of) all inductor storage loops
of each $\Phi$-DAC \cite{bunyk-architectural}. Table~\ref{t:energy} 
reports on\hyp{}chip energy dissipation values for various QPU sizes
and $\Phi$-DAC critical currents, showing that programming an example 
large\hyp{}scale device with 10~M qubits and 75~M couplers 
(15 per qubit \cite{topologies}) will dissipate only 
36~pJ on chip. With typical $\approx$30~$\mu$W cooling power available at the 20~mK QPU stage \cite{bluefors}, this accounts for 1.2~$\mu$s of QPU thermalization time.

The next step resets\fshyp{}initializes the qubits (\textit{cf.} \S\ref{s: algorithm}), 
during which each qubit transitions from a higher energy state to an 
intended ground state, generating spontaneous photon emissions, 
heating the QPU. Reed \textit{et al.} \cite{reed2010fast} 
demonstrate the suppression of these emissions 
using \emph{Purcell} filters, requiring 80~ns 
(120~ns) for 99\% (99.9\%) fidelity.

An $N_Q$ qubit, $N_C$ coupler, and five-bit precision QA device need to program a worst-case $5\cdot(N_Q + N_C)$ amount of data, which is 27~Kbytes for the current QA 
($N_Q$ = 5,436, $N_C$ = 37,440) and 100~Mbytes for a large-scale QA 
($N_Q$ = 10M, $N_C$ = 75M). Thus, to maintain today's microsecond level 
programming cycle time in future large-scale QA, programming control lines' 
bandwidth must be increased by a factor of $10^3$ (\textit{i.e.,} 
GHz bandwidth lines are needed). By Purcell filter design integration 
and sufficient amount of control line bandwidth, overall programming 
time (\textit{i.e.,} coefficient programming time + thermalization and reset) therefore reaches to 42~$\mu$s in a 10M-qubit large-scale QA device.

\subsection{Sampling}\label{s:sampling} The process of executing a QMI on a QA device is called \textit{sampling}, and the time taken for sampling is called the \textit{sampling time}. The sampling time is classified into three sub-components: the \textit{anneal}, \textit{readout}, and \textit{readout delay} times. A single QMI consists of multiple \textit{samples} of an input problem, with each sample annealed and read out once, followed by a readout delay (see Fig.~\ref{fig:timing}). Sampling a QMI begins after the QPU programming process.

\subsubsection{Anneal.} In this time interval, the QPU implements a QA algorithm (\S\ref{s: algorithm}) \cite{tech} to solve the input problem, where low-frequency \textit{annealing lines} control the annealing algorithm's schedule. The bandwidth of these control lines hence limits the minimum annealing time, which is one microsecond today. Weber \textit{et al.} \cite{lincolnlab} propose the use of flexible print cables with a moderate bandwidth ($\approx 100$~MHz) and high isolation ($\approx 50$~dB) for annealing, which potentially decrease annealing time to tens of nanoseconds.

\subsubsection{Readout} After annealing, the spin configuration of qubits
(\textit{i.e.,} the solution) is read out by measuring the qubits' persistent current ($I_p$) direction. This readout information propagates from the qubits to readout \textit{detectors} located at the perimeter of the QPU chip via \textit{flux bias lines}. Each flux bias line is a chain of electrical circuits called \textit{Quantum Flux Parametrons} (\textit{QFPs}), which detect and amplify qubits' $I_p$ to improve the readout signal-to-noise ratio. These QFP chains act like shift registers, propagating the information from qubits to detectors \cite{doi:10.1063/1.4939161}. In current QA devices with $N_Q$ qubits, there are $\sqrt{N_Q/2}$ flux bias lines, with each flux bias line responsible for reading out $\sqrt{2N_Q}$ qubits. Further, each flux bias line reads out one qubit at a time (\textit{i.e.,} time-division readout), thus a total of $\sqrt{N_Q/2}$ qubits are readout in parallel. Hence, the readout time depends on the qubits' physical~locations, the bandwidth of flux bias lines, and the signal integration time. For the current status of technology, the readout time is 25--150~$\mu$s per sample \cite{tech}. Nevertheless, recent research demonstrates promising fast readout techniques, which we describe next.

\begin{table}
\caption{The table shows the number of qubits read out in parallel by time-division (status quo) and frequency-multiplex (projected) readout schemes at various choices of QPU sizes and readout microresonator quality factors ($Q_r$).}
\label{t:readout}
\begin{small}
\begin{tabularx}{\linewidth}{X*{4}{c}}\toprule
\centering \multirow{3}{*}{\textbf{Qubits}} & \multicolumn{3}{c}{\textbf{Qubits readout in parallel}}\\ \cline{2-4} \noalign{\smallskip}
\centering & Time-division & \multicolumn{2}{c}{Frequency-multiplex} \\ \cline{3-4} \noalign{\smallskip} 
\centering & & $Q_r = 10^3$ \cite{doi:10.1063/1.4939161} & $Q_r = 10^6$ \cite{dorche2020high} \\
\bottomrule \noalign{\smallskip} 
\centering 512&  16 &  512 & 512\\
\centering 2,048& 32 & $\approx$ 666 & 2,048\\
\centering 5,436&  $\approx$ 52 & $\approx$ 666 & 5,436\\
\centering 10~M&  $\approx$ 2,200 & $\approx$ 666 & $\approx$ 666K\\
\bottomrule
\end{tabularx}
\end{small}
\end{table}

Chen \textit{et al.} \cite{doi:10.1063/1.4764940} and Heinsoo \textit{et al.} \cite{PhysRevApplied.10.034040} describe frequency-multiplex readout schemes that enable simultaneous readout of multiple qubits within a flux bias line. While there is no fundamental limit on the number of qubits read out simultaneously, a physical limit is imposed by the line width of qubits' readout microresonators and the 4--8~GHz operating band (6~GHz center frequency, 4~GHz bandwidth) of commercial microwave transmission line components used in the readout architecture \cite{doi:10.1063/1.4939161}. Microresonators with quality factor $Q_r$ can capture line widths up to 6/$Q_r$~GHz, thus enabling up to 4$\times Q_r$/6 qubits to be readout simultaneously. Table~\ref{t:readout} reports these results, showing that a $Q_r$ of $10^6$ will enable up to $\approx$666~K-qubit-parallel readout. This analysis assumes that each microresonator can be fabricated at exactly its design frequency, which is currently not the case. Further developments in understanding the RF properties of microresonators will therefore be needed to achieve this multiplexing performance.

In order to avoid sample-to-sample readout correlation, microresonators reading out the current sample's qubits must ring down before reading the next sample's qubits. McClure \textit{et al.} \cite{PhysRevApplied.5.011001} achieve ring-down times on the order of hundreds of nanoseconds by applying pulse sequences that rapidly extract residual photons exiting the microresonators after readout. Fast ring down can also be achieved by switching off the QFP (after the readout) coupled to a microresonator, and then switching on a different QFP that couples the microresonator to a lossy line. While QFP on-off switching takes hundreds of nanoseconds \cite{grover2020fast, 84613}, it ensures high fidelity readout.

Recent work by Grover \emph{et al.} \cite{grover2020fast} show the application of QFPs as isolators, achieving a readout fidelity of 98.6\% (99.6\%) in 80~ns (1~$\mu$s) only. Walter \emph{et al.} \cite{PhysRevApplied.7.054020} describe a single-shot readout scheme requiring only 48~ns (88~ns) to achieve a 98.25\% (99.2\%) readout fidelity. Their designs are also compatible with multiplexed architectures and earlier readout schemes, implying that by design integration readout time reaches on the order of microseconds per sample.

\subsubsection{Readout delay} After a sample's anneal-readout process, a \textit{readout delay} is added (see Fig.~\ref{fig:timing}). In this time interval, qubits are reset for next sample's anneal, and QA clients can specify times in the range 0--10~ms, and the default value is a conservative one millisecond. Nevertheless, about one microsecond is sufficient for high fidelity qubit reset (\S3.1)~\cite{reed2010fast}.

\subsection{Postprocessing}\label{s:post} This time interval is used for post-processing the solutions returned by QA for improving the solution quality~\cite{post}. Multiple samples' solutions are post-processed at once in parallel with the current QMI's annealer computation, whereas the final batch of post-processing occurs in parallel with the programming of next QMI (see Fig.~\ref{fig:timing}). Thus, the post-processing time does not factor into the overall processing time \cite{timing}.

\parabreak{}In summary, the projected overall programming time is 42~$\mu$s 
(programming: 4--40~$\mu$s, thermalization and reset: 2~$\mu$s), anneal time is one~$\mu$s\fshyp{}sample, readout time is one~$\mu$s\fshyp{}sample, and readout 
delay time is one~$\mu$s\fshyp{}sample. For a target sample count $N_{s}$, total QMI run time is 42 + 3$N_{s}$ $\mu$s.

\section{RAN Power Models and Cellular Targets}
\label{s:model}

This section describes power modeling in RANs (\S\ref{s:powermodels}), 4G/5G cellular computational targets (\S\ref{s:targets}), and QA qubit requirement to meet this computational demand (\S\ref{s:qubitcount}).

\begin{table*}
\caption{Table shows 4G and 5G cellular BBU computational targets in macro base stations operating at 64-QAM modulation and 0.5 coding rate. Time and frequency domain duty cycles are at 100\%. Values are in Tera operations per Second (TOPS).}
\label{t:targets}
\begin{small}
\begin{tabularx}{\linewidth}{X*{14}{r}}\toprule
\centering \multirow{2}{*}{\textbf{BBU Task}} & \multicolumn{1}{c}{\textbf{Reference}} & & \multicolumn{3}{c}{\textbf{4G} (B/W = 20 MHz)} & & \multicolumn{3}{c}{\textbf{5G} (B/W = 200 MHz)} & & \multicolumn{3}{c}{\textbf{5G} (B/W = 400 MHz)}\\ \cmidrule{2-2} \cmidrule{4-6} \cmidrule{8-10} \cmidrule{12-14}
\centering & $N_{A}$ = 1 & & $N_{A}$ = 2 & $N_{A}$ = 4 & $N_{A}$ = 8 & & $N_{A}$ = 32 & $N_{A}$ = 64 & $N_{A}$ = 128 & & $N_{A}$ = 32 & $N_{A}$ = 64 & $N_{A}$ = 128\\ \bottomrule \noalign{\smallskip} 
\centering DPD& 0.160 & & 0.320  & 0.640  & 1.280  & & 51.2  & 102.4  & 204.8 & & 102.4  & 204.8  & 409.6\\
\centering Filter& 0.400  & & 0.800  & 1.600  & 3.200  &  & 128.0  & 256.0 & 512.0 & & 256.0  & 512.0 & 1024.0\\
\centering FFT& 0.160  & & 0.320  & 0.640 & 1.280  & & 51.2 & 102.4 & 204.8 & & 102.4 & 204.8 & 409.6\\
\centering FD$_{\text{lin}}$& 0.090 & & 0.180  & 0.360  & 0.720   & & 28.8 & 57.6 & 115.2 & & 57.6 & 115.2 & 230.4\\
\centering FD$_{\text{nl}}$& 0.030 & & 0.120  & 0.480 & 1.920  &  & 307.2  & 1228.8 & 4915.2 & & 614.4  & 2457.6 & 9830.4\\
\centering FEC& 0.140 & & 0.140  & 0.280 & 0.560  &  & 22.4 & 44.8 & 89.6 & & 44.8 & 89.6 & 179.2\\
\centering CPRI & 0.720 & & 0.720  & 1.440  & 2.880   & & 115.2 & 230.4  & 460.8 & & 230.4 & 460.8  & 921.6\\
\centering PCP & 0.400 & & 0.800  & 1.600  & 3.200  &  & 12.8  & 25.6 & 51.2  & & 12.8  & 25.6 & 51.2\\
\centering \textbf{Total}& \textbf{2.100} & & \textbf{3.400}& \textbf{7.040} & \textbf{15.040} & &  \textbf{716.8} & \textbf{2,048.0} & \textbf{6,533.6} & &  \textbf{1,420.8} & \textbf{4,070.4} & \textbf{13,056.0}\\
\bottomrule
\end{tabularx}
\end{small}
\end{table*}

\subsection{Power Modeling}
\label{s:powermodels}
RAN power models account for power by splitting the BS or C-RAN functionality into the components and sub-components shown in Figs.~\ref{fig:cran}~and~\ref{fig:bs}. This section details these components and their associated power models. We follow the developments by Desset \textit{et al.} \cite{desset2012flexible} and Ge \textit{et al.} \cite{ge2017energy}.

\subsubsection{RAN Base Station}

\begin{figure}
\centering
\includegraphics[width=0.86\linewidth]{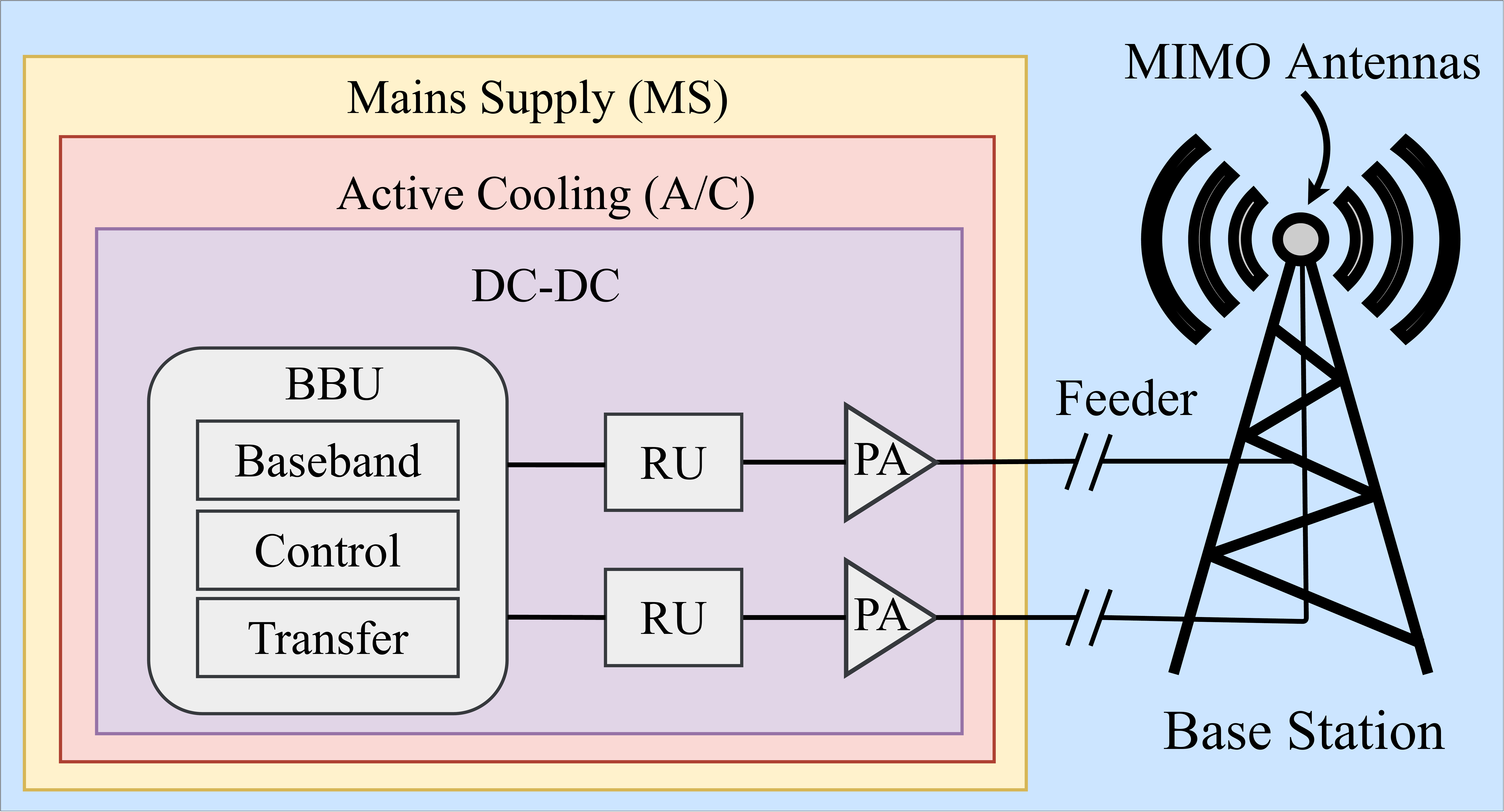}
\caption{A typical macrocell BS architecture.}
\label{fig:bs}
\end{figure}

A RAN BS (see Fig.~\ref{fig:bs}) is comprised of a baseband unit (BBU), a radio unit (RU), power amplifiers (PAs), antennas, and a power system (PS). The entire BS power consumption ($P_{\text{BS}}$) is then modeled as:
\begin{equation}
    P_{\text{BS}} = \frac{P_{\text{BBU}} + P_{\text{RU}} + P_{\text{PA}}}{(1-\sigma_{\text{A/C}})(1-\sigma_{\text{MS}})(1-\sigma_{\text{DC}})},
    \label{eq:powerBS}
\end{equation}

\noindent
where $P_i$ is the $i^{th}$ BS component's power consumption, and $\sigma_{\text{A/C}} (9\%)$, $\sigma_{\text{MS}} (7\%)$, and $\sigma_{\text{DC}}(6\%)$ correspond to fractional losses of Active Cooling (A/C), Mains Supply (MS), and DC--DC conversions of the power system respectively \cite{ge2017energy}.

The BBU performs the processing associated with digital baseband (BB), and control and transfer systems. The baseband includes computational tasks such as digital pre-distortion (DPD), up/down sampling or filtering, OFDM-FFT processing, frequency domain (FD) mapping/demapping and equalization, and forward error correction (FEC). The control system undertakes the platform control processing (PCP), and the transfer system processes the eCPRI transport layer. The total BBU power consumption ($P_{\text{BBU}}$) is then \cite{desset2012flexible, ge2017energy}: 
\begin{multline}
P_{\text{BBU}} = P_{\text{DPD}} + P_{\text{Filter}} + P_{\text{FFT}} + P_{\text{FD}_{\text{lin}}} + P_{\text{FD}_{\text{nl}}} + P_{\text{FEC}}\\ + P_{\text{PCP}} + P_{\text{CPRI}} + P_{\text{Leak}},
\label{eq:powerDU}
\end{multline}

\noindent
where $P_i$ is the $i^{th}$ computational task's power consumption, and $P_{\text{Leak}}$ is the leakage power resulted from the employed hardware in processing these baseband tasks. FD processing is split into two parts, with linear and non-linear scaling over number of antennas \cite{desset2012flexible, ge2017energy}. The RU performs analog RF signal processing, consisting of clock generation, low-noise and variable gain amplification, IQ modulation, mixing, buffering, pre-driving, and analog--digital conversions. RU power consumption ($P_{\text{RU}}$) scales linearly with number of transceiver chains, and each chain consumes about 10.8~W power \cite{desset2012flexible}. For macro-cell BSs, each PA (including antenna feeder) is typically configured at 102.6~W power consumption \cite{ge2017energy}.

\subsubsection{C-RAN} In the C-RAN architecture, BS processing functionality is amortized and shared, where Remote Radio Heads (RRHs) perform analog RF signal processing and a BBU-pool performs digital baseband computation (of many BSs) at a centralized datacenter (see Fig.~\ref{fig:cran}). Fronthaul (FH) links connect RRHs with the centralized BBU-pool. To relax the FH latency and bandwidth requirements, a part of baseband computation is performed at RRH sites. Several such split models have been proposed \cite{8479363, 3gpp}. We consider a split where RRHs perform low Layer 1 baseband processing, such as cyclic prefix removal and FFT-specific computation. The power consumption of C-RAN ($P_{\text{C-RAN}}$) is then:
\begin{equation}
    P_{\text{C-RAN}} = P_{\text{BBU}} + P_{\text{PS}_{\text{BBU}}} + \sum_{k=1}^{N_{RRH}} \Big\{P_{\text{RRH}_k} + P_{\text{PS}_{\text{RRH}_k}} + P_{\text{FH}_k}\Big\},
    \label{eq:powerCRAN}
\end{equation}

\noindent
where $P_{k}$ is the $k^{th}$ C-RAN component's power consumption and $N_{RRH}$ is the number of RRHs. Fronthaul power consumption depends on the technology, and for fiber-based ethernet or passive optical networks, it can be modeled by assuming a set of parallel communication channels as \cite{7437385, alimi2019energy}:
\begin{equation}
    P_{\text{FH}_k} = \rho_kR_{\text{FH}_{k}}, \hspace{10pt} \rho_k = P_{\text{FH}_{k, \text{max}}}/{C_{\text{FH}_k}}
    \label{eq:powerFH}
\end{equation}

\noindent
where $\rho_k$ is a constant scaling factor, $R_{\text{FH}_k}$ and $C_{\text{FH}_k}$ represent the traffic load and the capacity of the $k^{th}$ fronthaul link respectively. For a link capacity of 500~Mbps, $P_{\text{FH}_{k, \text{max}}}$ is typically \textit{ca.} 37 Watts \cite{liu2018designing}.

\subsection{Cellular Processing Requirements}
\label{s:targets}

This section describes 4G/5G cellular computational targets in estimated Tera operations per Second (TOPS) the BBU needs to process, and it depends on parameters such as the bandwidth (B/W), modulation (M), coding rate (R), number of antennas ($N_A$), and time (\textit{dt}) and frequency (\textit{df}) domain duty cycles. Prior work \cite{desset2012flexible, ge2017energy} present these TOPS complexity values for individual BBU tasks in a reference scenario (B/W = 20~MHz, M = 6, R = 1, $N_A$ = 1, \textit{dt} = \textit{df} = 100\%), which we replicate in Table~\ref{t:targets} as Reference. The scaling of these values follow \cite{desset2012flexible, ge2017energy}:
\begin{equation}
    \text{TOPS}_{\text{target}} = \text{TOPS}_{\text{ref}}\prod_{k}\Bigg(\frac{X_{\text{target}}}{X_{\text{ref}}}\Bigg)^{s_k}
    \label{eq:tops}
\end{equation}

\noindent
where X $\in$\big\{B/W, M, R, $N_A$, \textit{dt}, \textit{df}\big\} and \textit{k}$\in$ [1,6] respectively. The scaling exponents \{$s_1$, $s_2$, $s_3$, $s_4$, $s_5$, $s_6$\} are \{1,0,0,1,1,0\} for DPD, Filter, and FFT, \{1,0,0,1,1,1\} for FD$_{\text{lin}}$, \{1,0,0,2,1,1\} for FD$_{\text{nl}}$, \{1,1,1,1,1,1\} for CPRI and FEC, and \{0,0,0,1,0,0\} for PCP \cite{desset2012flexible, ge2017energy}. These exponents are determined based on the dependence of BBU operation with the corresponding parameters \cite{desset2012flexible, ge2017energy}. Table~\ref{t:targets} reports the TOPS complexity values for representative 4G and 5G scenarios.

\subsection{QA Qubit Count Requirements}
\label{s:qubitcount}

This section describes our approach in estimating the QA qubit requirement that meet the 4G/5G cellular baseband computational demand (\S\ref{s:targets}). To compute this, we convert the target TOPS (Table~\ref{t:targets}) into target \textit{problems per second (PPS)}, then estimate the number of qubits QA requires to achieve this PPS, individually for baseband computational tasks. We formulate it as:
\begin{align}
    N_Q = \sum_k N_{Q,k}\\
    N_{Q,k} = PPS_k\times N_{Q,p,k} \times T_{p,k} \label{eq:qubits}\\
    PPS_k = TOPS_k/\text{Number of operations per problem}
\end{align}

\noindent
where $N_Q$ is the total number of qubits the QA requires for the entire baseband processing, and $N_{Q,k}$ is the qubit requirement for the $k^{th}$ baseband task. $PPS_k$ is the target problems per second, $N_{Q,p,k}$ is the number of qubits per problem, and $T_{p,k}$ is the run time per problem, of the $k^{th}$ baseband task. We next demonstrate how to compute these values for $\text{FD}_{\text{nl}}$ and FEC tasks with running examples.

\paragraph{$\textbf{FD}_{\textbf{nl}}$ Qubit Requirement.} The $\text{FD}_{\text{nl}}$ task corresponds to the MIMO detection problem \cite{albreem2019massive} whose objective is to demodulate the received soft symbols into bits. Solving an $\text{FD}_{\text{nl}}$ problem requires on average 80~$\times$~($Z$/64)$^\text{2}$ million operations for a $Z\times Z$ (\textit{Z}-users, \textit{Z}-antennas) system\footnote{A $64\times 64$ MIMO detection problem requires 80 million operations \cite{jalden2004maximum}, and it scales quadratic with number of antennas \cite{desset2012flexible, ge2017energy}.} via state-of-the-art \textit{Sphere Decoding} algorithm \cite{jalden2004maximum}. Solving the same problem using QA requires $N_{\text{bps}}\times Z$ qubits, where $N_{\text{bps}}$ is the number of bits per symbol in the modulation scheme (see \cite{kim2019leveraging} for full derivation). Thus for a typical 5G scenario: $64\times 64$ MIMO system with 64-QAM modulation (\textit{i.e.,} six bits per symbol), $PPS_{\text{FD}_{\text{nl}}}$ is 30.72M (= 2457.6 TOPS/80M operations), $N_{Q,p,\text{FD}_{\text{nl}}}$ is 384 qubits, and $T_{p,\text{FD}_{\text{nl}}}$ is 42 + 3$N_s$ $\mu$s (\S\ref{s:latency}). Substituting these values in Eq.~\ref{eq:qubits} shows that the 5G $\text{FD}_{\text{nl}}$ processing requires 1.2M qubits with $N_s = 20$ samples.

\paragraph{FEC Qubit Requirement.} The FEC task corresponds to channel decoding that aims to correct the bit errors that interference and vagaries of the wireless channel inevitably introduce into the user data. We consider Low Density Parity Check (LDPC) codes employed in the 5G-NR traffic channel for FEC evaluation \cite{3gppcoding}. Decoding an \textit{(M, N)}-LDPC code with average row weight $w_r$ and column weight $w_c$ in its parity check matrix via state-of-the-art \textit{belief propagation} algorithm requires $N + 3w_r^2M - w_rM + 2w_c^2N + 4w_cN$ operations per iteration \cite{fernandes2010parallel}, where \textit{M} and \textit{N} are the number of rows and columns in the LDPC parity check matrix respectively. Solving the same problem using QA requires $N + Mt$ qubits, where $t = \text{arg}\min_{n \in \mathbb{Z}} \{2^{n+1} - 2 \geq w_r - (w_r\mod 2)\}$---see \cite{10.1145/3372224.3419207} for full derivation. Thus for the 5G's longest LDPC code with base-graph-1 parity check matrix ($M$ = 4224, $N$ = 8448, $w_r$ = 8.64, $w_c$ = 20) \cite{3gpp}, $PPS_{\text{FEC}}$ is 600K (= 89.6 TOPS/150M operations)---for typical 20 decoding iterations, $N_{Q,p,FEC}$ is 21,120 qubits, and and $T_{p, FEC}$ is $80 + 3N_s$ $\mu$s (\S\ref{s:latency}). Substituting these values in Eq.~\ref{eq:qubits} shows that the 5G FEC processing requires 1.29M qubits with $N_s = 20$ samples.
\begin{table}
\caption{QA qubit requirement at various problem run times to achieve spectral efficiency equal to CMOS processing, in a 5G BS scenario with 400~MHz BW and 64 antennas.}
\label{t:qubits-time}
\begin{small}
\begin{tabularx}{\linewidth}{X*{5}{c}}\toprule
\centering $N_s$: & 1 & 20 & 50 & 100\\ \cline{1-5} \noalign{\smallskip}
\centering $T_{p,k} (\mu s):$ & 45 & 102 & 192 & 342 \\
\midrule \noalign{\smallskip} 
\centering $N_{Q, FD_{nl}}:$& 530K & 1.20M & 2.26M & 4.03M \\
\centering $N_{Q, FEC}$:& 570K & 1.29M & 2.43M & 4.34M\\
\centering $N_Q:$& 1.60M & 1.99M & 6.25M & 11.16M\\
\bottomrule
\end{tabularx}
\end{small}
\end{table}

\begin{figure*}[ht]
\centering
\begin{subfigure}[b]{0.32\linewidth}\centering
\includegraphics[width=\linewidth]{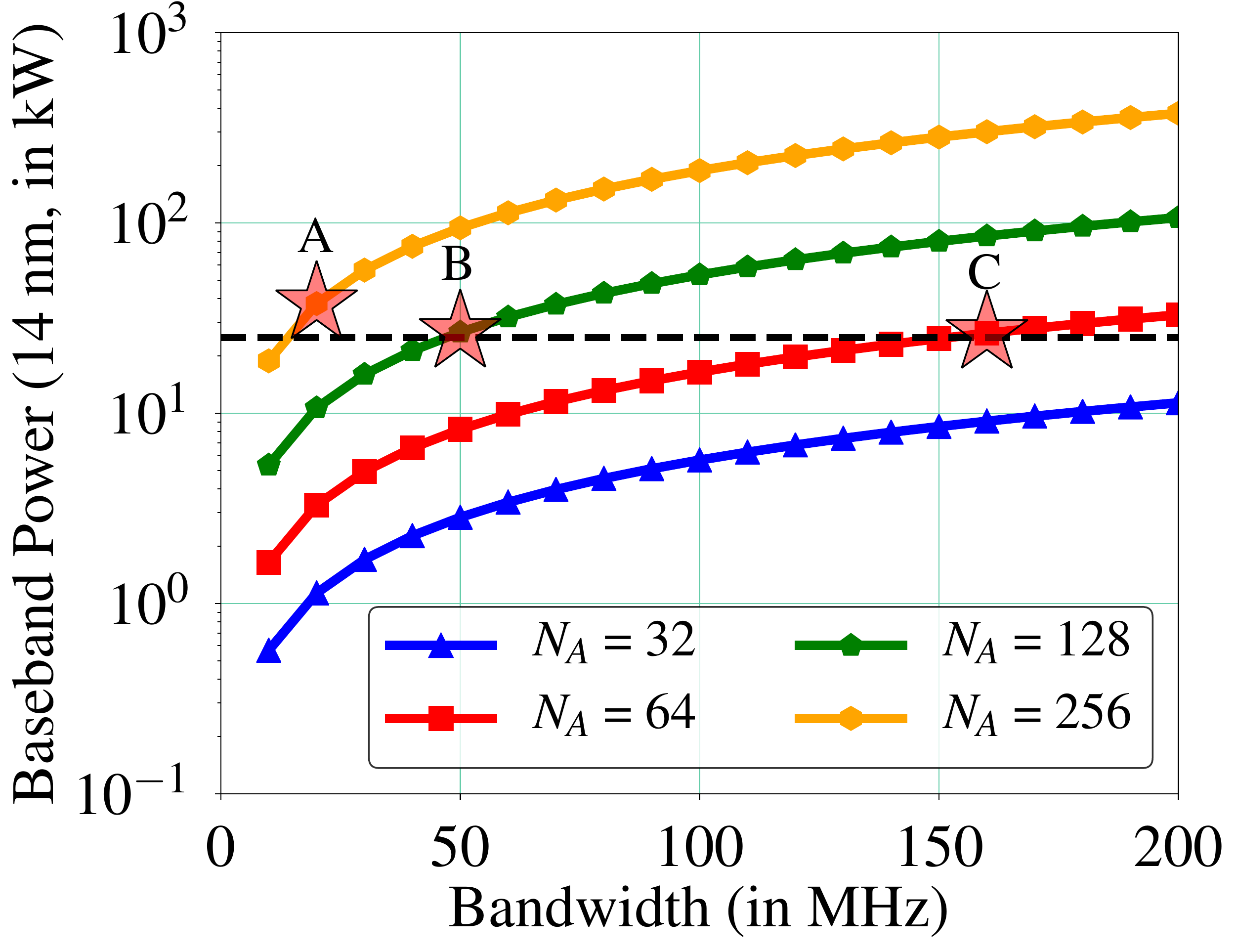}
\caption{$P_{\text{14 nm CMOS}}$ vs $P_{\text{QA}}$.}
\label{fig:14nm}
\end{subfigure}
\hfill
\begin{subfigure}[b]{0.32\linewidth}\centering
\includegraphics[width=\linewidth]{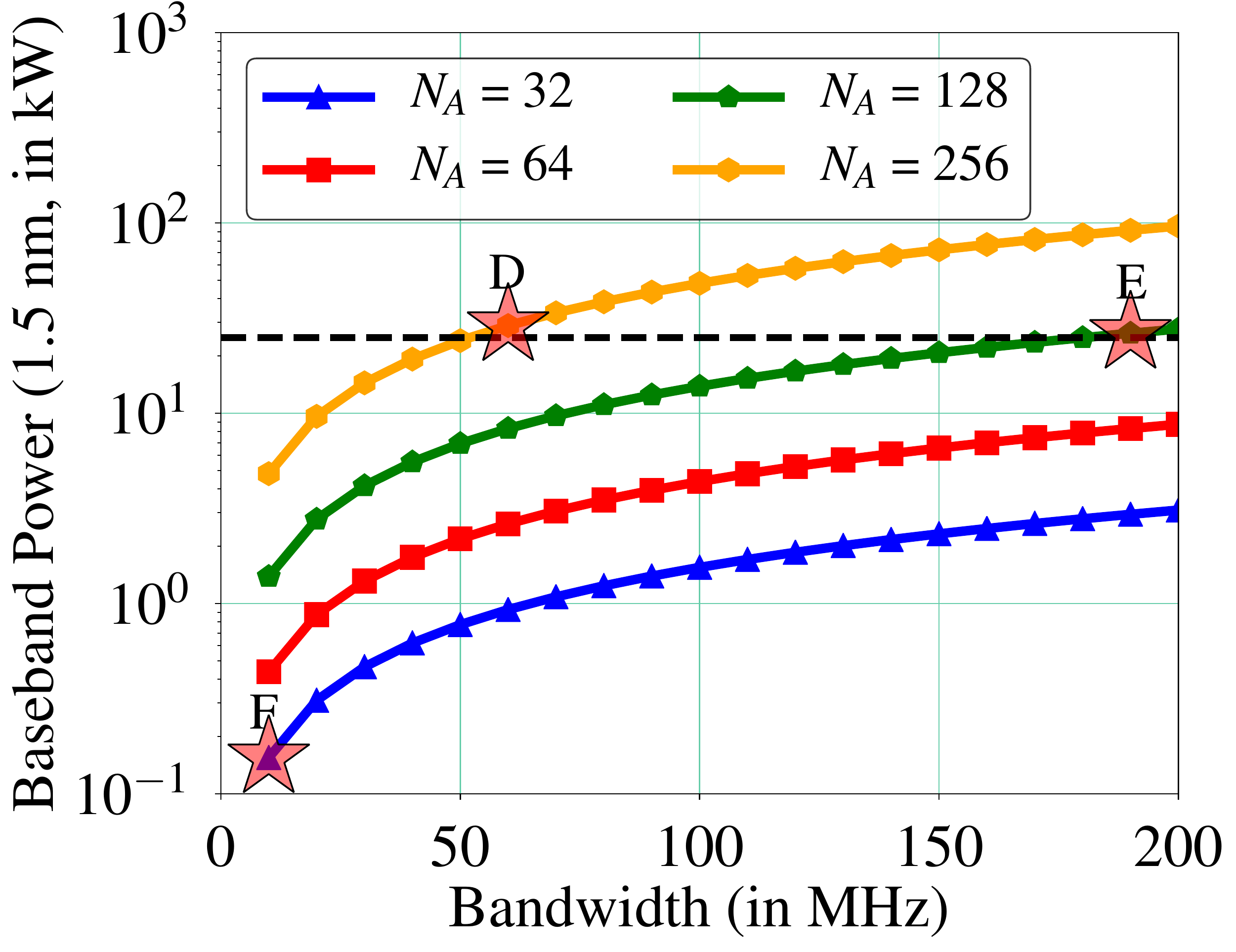}
\caption{$P_{\text{1.5 nm CMOS}}$ vs $P_{\text{QA}}$.}
\label{fig:1.5nm}
\end{subfigure}
\hfill
\begin{subfigure}[b]{0.35\linewidth}\centering
\includegraphics[width=\linewidth]{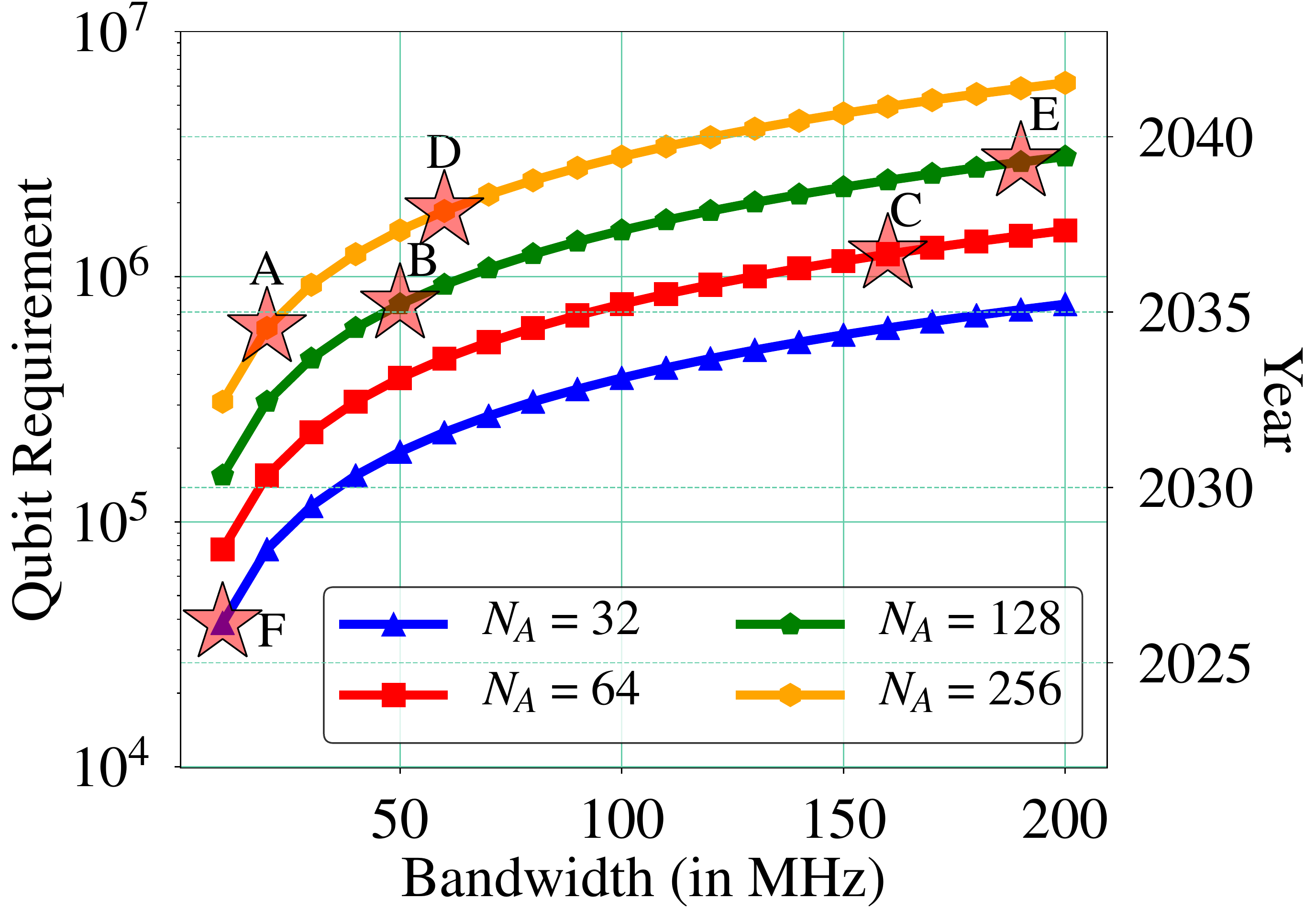}
\caption{QA qubit requirement.}
\label{fig:qubits}
\end{subfigure}
\hfill
\caption{Power consumption of BBU's baseband and its associated power system using \textbf{(a)} 14 nm CMOS and (\textbf{b}) 1.5 nm CMOS hardware in various base station operation scenarios in the 5G frequency range \cite{3gpp}. The dotted horizontal line in (\textbf{a}) and (\textbf{b}) is the QA power consumption of 25~kW. \textbf{(c)} The number of qubits QA requires to match the spectral efficiency of CMOS in the same scenarios. Points A--E respectively show the smallest bandwidth at which QA benefits in power over CMOS at each antenna count, and Point F shows the smallest practically feasible scenario QA enables with 39K qubits (see \S\ref{s:timelines}).}
\label{fig:discussion}
\end{figure*}

5G's $\text{FD}_{\text{nl}}$ and FEC tasks correspond to 75\% of the baseband computation load. For the remaining 25\% of baseband computational load, we project a proportionate number of qubits for their respective processing requirements.  Table~\ref{t:qubits-time} reports the number of qubits the QA requires as a function of problem run time ($T_{p,k}$), showing that with $T_{p,k}$ of \{45, 102, 192, 342\}~$\mu$s, QA requires \{1.6, 1.99, 6.25, 11.16\} million qubits respectively to satisfy the 5G baseband computational demand. The number of samples ($N_s$) represent the required QA target fidelity in terms of error performance---when $N_s$ is 20, QA must reach ground state of the input problem in 20 anneals. Hence, QA must meet these $T_{p,k}$ and $N_Q$ combinations to achieve spectral efficiency equal to CMOS processing in 5G wireless networks. While we demonstrate an example scenario with 400~MHz BW, 64-antennas, 64-QAM modulation, and 0.5 coding rate, a similar methodology can be applied to estimate network-specific qubit requirements. Figure~\ref{fig:qubits} shows this qubit requirement for various bandwidths and antenna choices at 102~$\mu$s problem run time.

\section{Power and Cost Comparison}
\label{s:results}

Our methodology compares CMOS and QA processing 
at equal spectral efficiency outcomes. We 
specify the same BBU targets (Table~\ref{t:targets}) with CMOS 
and QA hardware, ensuring equal bits processed per second per Hz per km$^2$.

\begin{figure*}[ht]
\centering
\begin{subfigure}[b]{0.49\linewidth}
\includegraphics[width=\linewidth]{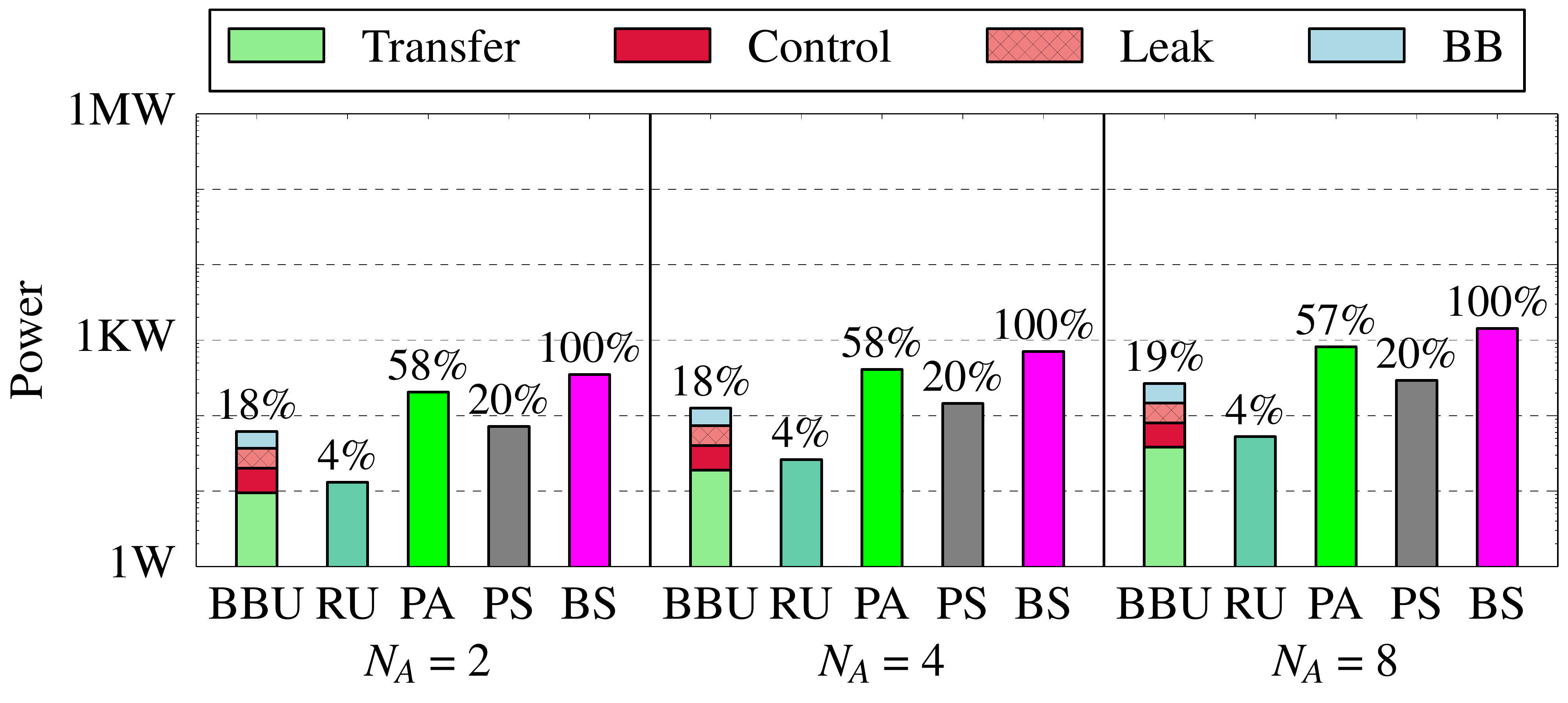}
\caption{A 4G scenario with 20~MHz bandwidth.}
\label{fig:4g-silicon}
\end{subfigure}
\hfill
\begin{subfigure}[b]{0.49\linewidth}
\includegraphics[width=\linewidth]{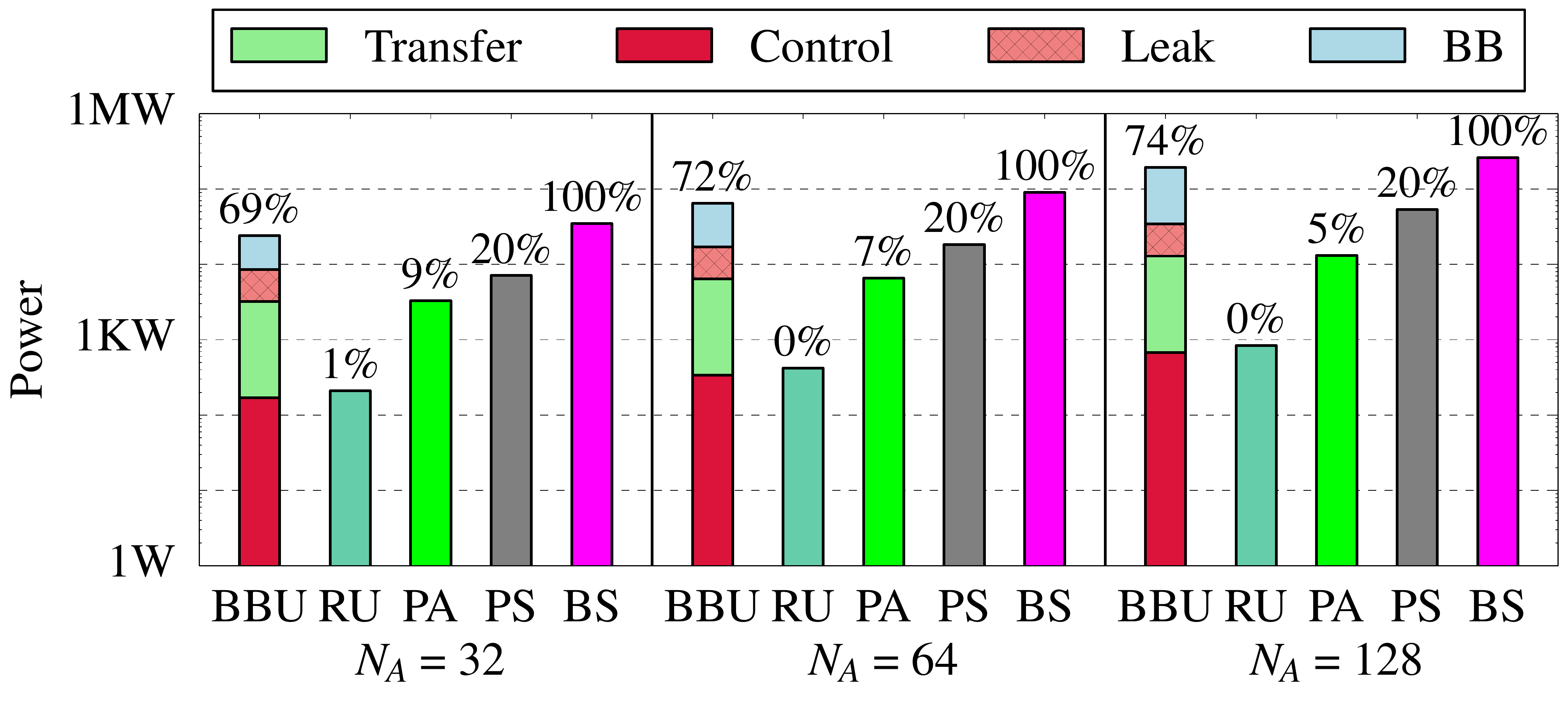}
\caption{A representative 5G scenario with 400~MHz bandwidth.}
\label{fig:5g-silicon}
\end{subfigure}
\hfill
\caption{Power consumption of silicon 14~nm CMOS processing in 4G and 5G base stations. BBU bar plots are shown with its sub-components (see legend, \S4.1.1) in increasing order of power consumption from bottom to top. The percentages (rounded to nearest integer) show the power contribution of that particular BS component (labeled on X-axis) to the total BS power. The BS power at $N_A$=\{2, 4, 8, 32, 64, 128\} is \{0.35, 0.71, 1.43, 34.7, 89.9, 261.3\}~kW, in their respective scenarios.}
\label{fig:4g-5g-bs}
\end{figure*}

\begin{figure*}[ht]
\centering
\begin{subfigure}[b]{0.49\linewidth}\centering
\includegraphics[width=\linewidth]{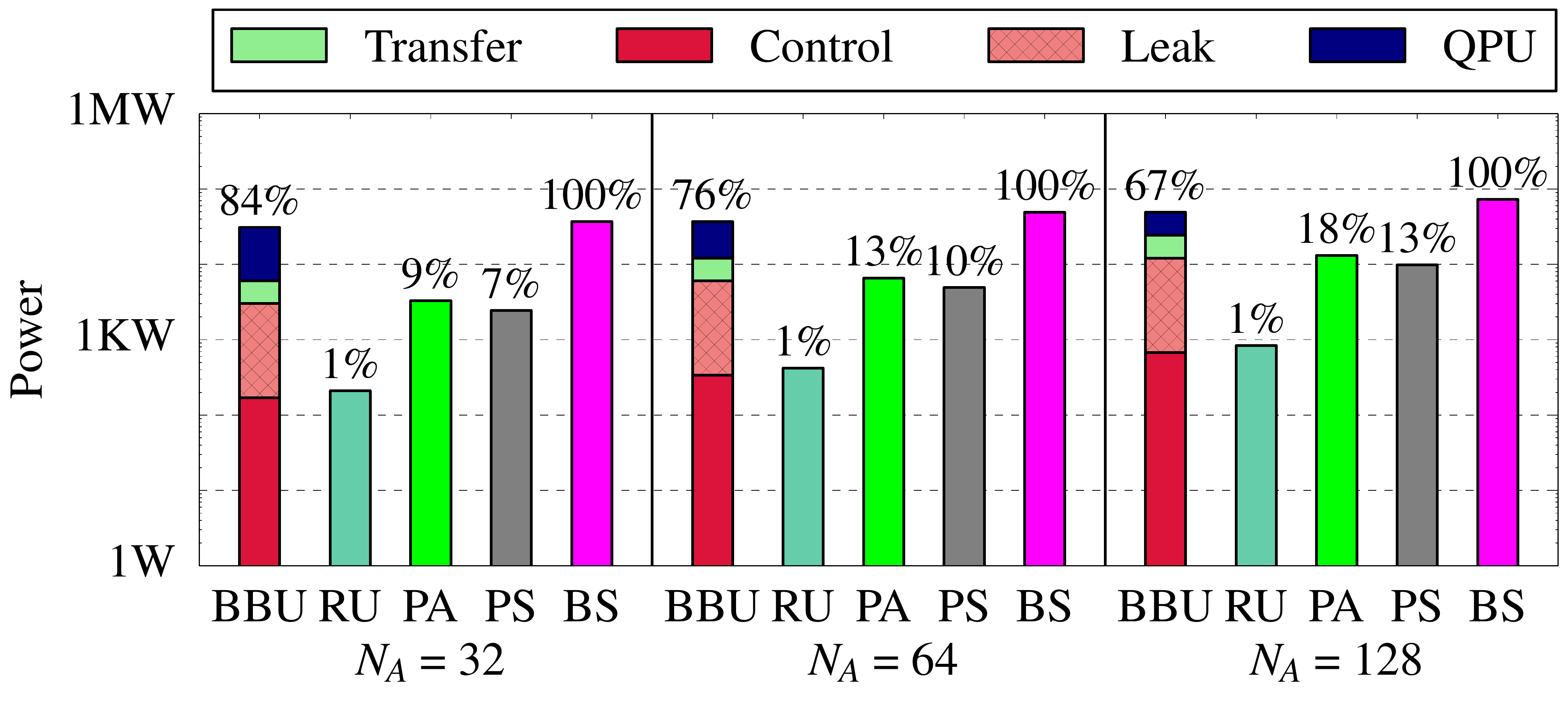}
\caption{A representative 5G BS scenario with 400~MHz bandwidth.}
\label{fig:qa-5g}
\end{subfigure}
\hfill
\begin{subfigure}[b]{0.49\linewidth}\centering
\includegraphics[width=\linewidth]{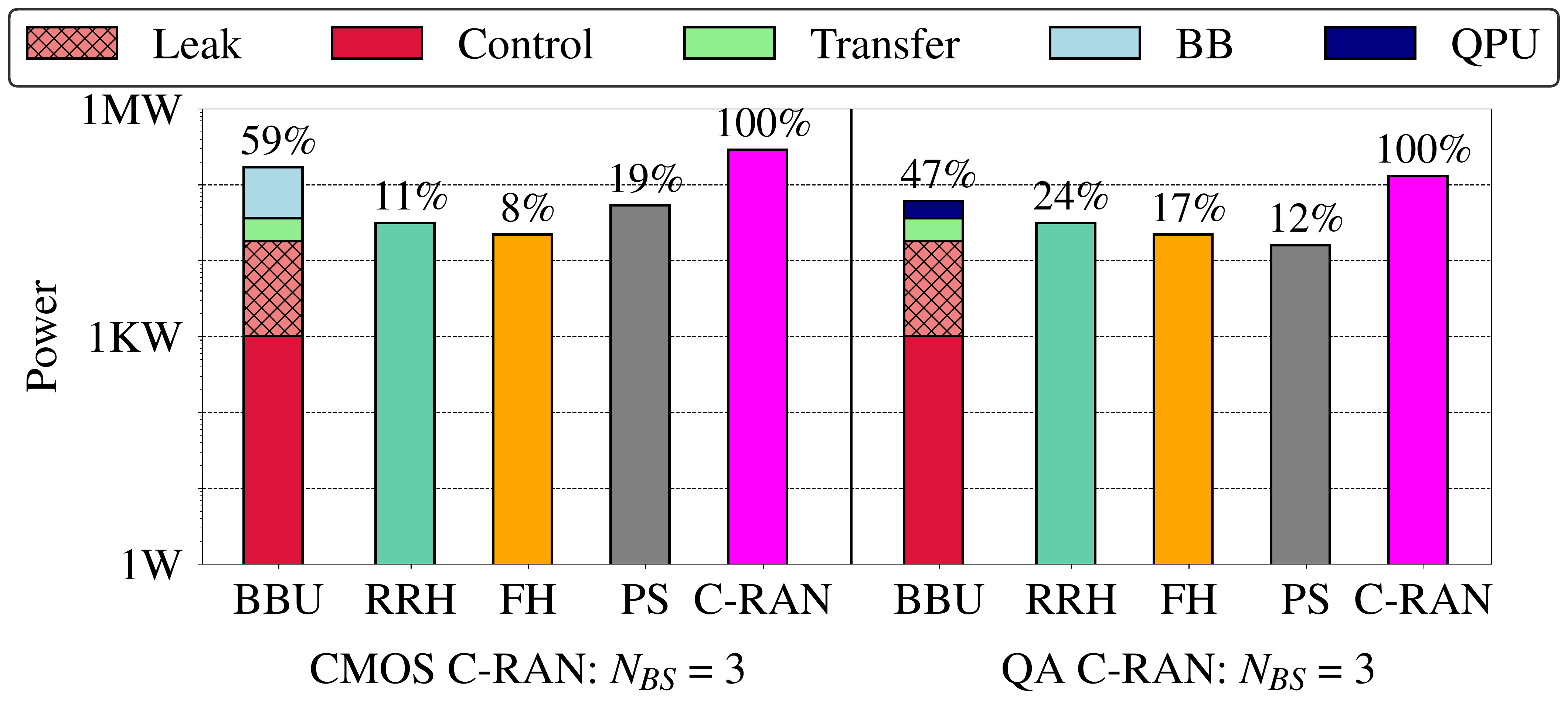}
\caption{A C-RAN with three 400~MHz-bandwidth 64-antenna BSs.}
\label{fig:qa-cran}
\end{subfigure}
\hfill
\caption{\textbf{(a)} Power consumption of a 5G BS where QA is used of the BBU's baseband processing. The BS power at $N_A$ = \{32, 64, 128\} is \{37, 49, 73\}~kW respectively. \textbf{(b)} Power consumption of CMOS (290~kW) and QA (131~kW) processing in C-RAN scenario with three base stations. In both \textbf{(a)} and \textbf{(b)}, BBU's further computation (\textit{i.e.,} Control and Transfer systems) is processed by 14~nm CMOS silicon.  BBU bar plots are shown with its sub-components (see legend, \S4.1.1) in increasing order of power from bottom to top. The percentages (rounded to nearest integer) correspond to components labeled on X-axis.}
\label{fig:qa-bs-results}
\end{figure*}

Power consumption of CMOS hardware depends on its \textit{performance-per-watt} efficiency and the amount of computation at hand. Technology scaling improves this efficiency from generation to generation, inversely proportional to the square of its transistors' core supply voltage ($V_{dd}$) \cite{stillmaker2017scaling}. A 65~nm CMOS device ($V_{dd}$ = 1.1 V) has a 0.04 TOPS/Watt efficiency, from which we compute the same for today's 14~nm CMOS ($V_{dd}$ = 0.8 V) and future 1.5~nm CMOS ($V_{dd}$~=~0.4 V), via $V_{dd}^2$ scaling, and they obtain a 0.076 and 0.3 TOPS/Watt efficiency respectively \cite{desset2012flexible, itrs, itrs_old}. Using this hardware efficiency and the TOPS requirements of Table~\ref{t:targets}, we compute CMOS hardware power consumption. Additional power results from leakage currents in CMOS transistor channel, and this leakage power is set to 30\% of dynamic power \cite{desset2012flexible}.

Power consumption of D-Wave's QA is \textit{ca.} 25~kW, dominated by its refrigeration unit (see Supplementary information--\cite{king-naturecomms2021}). Additional power draw due to the computation at hand is fairly negligible compared to the QA refrigeration power, since the QPU resources used for computation are thermally isolated in a superconducting environment. This power requirement is further not expected to significantly scale up with increased qubit numbers \cite{villalonga2020establishing, king-naturecomms2021}, due to the fairly constant power consumption of pulse-tube dilution refrigerators which are used to cool the QPU in practice \cite{nextdwave, king-naturecomms2021, bluefors}. More general NISQ processors such as Google's Sycamore (see Supplementary information--\cite{google-nature19}) and IBM's Rochester \cite{ibmrochester} also show a similar \textit{ca.} 25~kW power consumption and a fairly constant scaling with increased qubit numbers \cite{villalonga2020establishing}. However, to maintain this 25~kW power for the entire 5G baseband processing, sufficient amount of qubits are required, all under the same refrigeration unit. This raises the question---how many qubits are allowed in a QA refrigeration unit?

To answer this question, we consider the physical size of qubits in their unit cell packaging (a \emph{die}) versus the available space in the dilution refrigerator. The number of useful square dies ($N_d$) of length $L_d$ placed onto a \emph{wafer} of radius $R_w$ is approximately \cite{de2005investigation}: $N_{d} = \frac{\pi R_w^2}{L_d^2} - \frac{1.16\pi R_w}{L_d}$. A square die of eight qubits requires 335$\times$335 $\mu m^2$ QPU chip area with $L_d$ = 335 $\mu$m \cite{bunyk-architectural}, and a dilution refrigerator's experimental space has a radius $R_w$ = 250 \textit{mm} \cite{bluefors}. Substituting these values in the above equation gives $N_d$ $\approx$ 1.75M, which implies $\approx$14 million qubits allowed in a refrigeration unit. Since qubit count estimates for 5G (\textit{cf.} \S\ref{s:qubitcount}, \S\ref{s:timelines}) are well below this allowed limit, QA power consumption is 25~kW for 5G baseband~processing.

\paragraph{Results and discussion:} Applying the foregoing power analysis, 
Fig.~\ref{fig:4g-5g-bs} reports power consumption results of 4G and 5G BSs with 14~nm CMOS hardware. In Fig.~\ref{fig:4g-silicon}, we see that the power amplifier (PA) is the dominating component of 4G BS power consumption, as identified in several prior works \cite{desset2012flexible, ge2017energy, alimi2019energy}, accounting for 57--58\% of the total BS power. But, as the network scales to higher bandwidth and antennas envisioned in 5G, the BBU becomes the dominant power consuming component (see Fig.~\ref{fig:5g-silicon}), accounting for 69--74\% of the total BS power. This quick escalation in power from 0.35--1.43~kW in 4G to 34.7--261.3~kW in 5G is mainly due to the quadratic dependency of FD processing with number of antennas (\S\ref{s:powermodels}), and the increased network bandwidth consequence of millimeter-wave communication.

Fig.~\ref{fig:qa-5g} reports the power consumption results of 5G BS, where QA is used for BBU's baseband processing. In comparison to CMOS--Fig.~\ref{fig:5g-silicon}, QA reduces BS power by 41~kW and 188~kW in 64 and 128 antenna systems. Fig.~\ref{fig:qa-cran} shows power consumption in a C-RAN setting with three 64-antenna BSs, where the fronthaul is allowed a 100~Gbps bandwidth. In comparison to CMOS, QA processing reduces C-RAN power by 159~kW (55\% lower). Table~\ref{t:costsavings} reports the OpEx cost savings and carbon emission reductions associated with the respective power savings, computed by considering an average \$0.143 (USD) electricity price and 0.92 pounds of $CO_2$ equivalent emitted per kWh~\cite{blscost, carbon}. To provide economic benefit over CMOS hardware, assuming CMOS CapEx is negligible, future QAs' CapEx must be lower than the respective OpEx savings. For instance, if QA was to be employed in a C-RAN scenario, a CapEx lower than 200K, 400K, 1M, and 2M USD will provide economic benefit over CMOS in one, two, five, and 10 years, respectively.

\begin{table}
\caption{Summary of OpEx electricity cost savings (in USD) and $CO_2$ emissions reduction (in metric kilotons) QA will achieve in comparison to CMOS in 5G network scenarios. The number of antennas in C-RAN BSs is $N_A$ = 64.}
\label{t:costsavings}
\begin{small}
\begin{tabularx}{\linewidth}{X*{7}{c}}\toprule
\centering \multirow{3}{*}{\textbf{Years}} & \multicolumn{2}{c}{BS ($N_A$ = 64)} & \multicolumn{2}{c}{BS ($N_A$ = 128)} & \multicolumn{2}{c}{C-RAN} \\ \cline{2-7} \noalign{\smallskip}
\centering & Cost (\$) & $CO_2$ (kt) & Cost & $CO_2$ & Cost & $CO_2$  \\
\cmidrule{1-7}
\centering 1 & 50K & 0.15 & 235K & 0.68 & 200K & 0.57 \\
\centering 2 & 100K & 0.30 & 471K & 1.37 & 400K & 1.15\\
\centering 5 & 250K & 0.75 & 1.17M & 3.43 & 1M & 2.87\\
\centering 10 & 500K & 1.50 & 2.35M & 6.87 & 2M & 5.75\\
\bottomrule
\end{tabularx}
\end{small}
\end{table}

\section{Feasibility Timeline and Discussion}
\label{s:timelines}

This section presents our projected QA feasibility timeline, describing year-by-year milestones on the application of QA to wireless networks. Our approach is to compare power consumption of QA and CMOS in various base station scenarios, then compute the QA qubit requirement to equal spectral efficiency to CMOS in the same scenarios. We next project the year by which these qubit numbers become available in the QA hardware by extrapolating the historical QA qubit growth trend into future. Figures \ref{fig:timeline} and \ref{fig:discussion} report these results.

\parahead{Roadmap for feasibility.} The processing of a base station with 10-MHz bandwidth and 32 antennas (Point `F' in Fig.~\ref{fig:qubits}) requires 39K qubits in the QA hardware for QA to equal spectral efficiency to CMOS, and this qubit requirement is projected to become available by the year 2026 (Figs.~\ref{fig:timeline}, \ref{fig:qubits}). However, leveraging QA for such a system leads to increased power consumption in comparison to both 14~nm and 1.5~nm CMOS devices (Figs.~\ref{fig:14nm}, \ref{fig:1.5nm}).

\parahead{Roadmap for Power dominance.} From Figs.~\ref{fig:14nm} and \ref{fig:1.5nm}, we see that for a given antenna count, the lowest bandwidth for which QA achieves power advantage over 14~nm CMOS are 20~MHz bandwidth 256-antenna (Point `A'), 50~MHz bandwidth 128-antenna (Point `B'), and 160~MHz bandwidth 64-antenna (Point `C') systems. In comparison to 1.5~nm CMOS, such points correspond to 60~MHz bandwidth 256-antenna (Point `D'), and 190~MHz bandwidth 128-antenna (Point `E') systems. Fig.~\ref{fig:qubits} shows the number of qubits required in the QA hardware to process these systems (Points A--E) with equal spectral efficiency to CMOS. The figure shows that to achieve a power dominance over 14~nm CMOS, at least 618K qubits (Point `A') are required in the QA hardware, and this qubit requirement is projected to become available by the year 2035 (Figs.~\ref{fig:timeline}, \ref{fig:qubits}). QA with at least 1.85M qubits benefit in power over 1.5~nm CMOS, and such a QA is predicted to become available by the year 2038 (Figs.~\ref{fig:timeline}, \ref{fig:qubits}). In summary, our analyses show that power advantage of QA over CMOS is a predicted 14--17 years away. Fig.~\ref{fig:timeline} summarizes Fig.~\ref{fig:discussion} in a feasibility timeline, showing the years by which QA enables these base station operation scenarios along with their associated power advantage/loss.

\section{Conclusion}
\label{s:conclusion}

While the conventional assumption that 
CMOS hardware will achieve nextG cellular processing targets may
well hold true, this paper
makes the case for the possible future feasibility and 
potential power advantage of QA over CMOS. 
Our extensive analysis of current QA technology projects
quantitative targets that future QAs may well meet in order to 
provide benefits over CMOS in terms of performance, power, and cost. 
While we acknowledge the practical deployment of quantum processors 
to be at least tens of years away, this early study informs
future quantum hardware design and RAN architecture evolution. Furthermore, fundamental physical advances in the QA technology itself, which we do not leverage in the projections given in this paper, may offer even further benefits, advantaging our
projected timelines. Examples of these advances include faster annealing times ($< 40$~ns) and/or qubits with longer coherence lifetimes (such as the qubits in IARPA's QEO and DARPA's QAFS QA chips \cite{darpa}) that enable coherent quantum annealing regimes, benefiting future QA spectral efficiency \cite{fastanneal, yan2016flux}.

\section*{Acknowledgements}
This research is supported by National Science Foundation (NSF) Award CNS-1824357. We thank Andrew J. Berkley, Keith Briggs, Andrew D. King, Catherine McGeoch, Davide Venturelli, and Catherine White for useful discussions. P.A.W. is supported by the Engineering and Physical Sciences Research Council (EPSRC) Hub in Quantum Computing and Simulation, Grant Ref.~EP/T001062/1.

\bibliographystyle{concise2}
\bibliography{reference}

\end{document}